\documentclass[compsoc, conference, a4paper, 10pt, times]{IEEEtran}

\usepackage{graphicx}
\usepackage{caption}
\usepackage{rotating}
\usepackage{latexsym}
\usepackage{amssymb,amsthm,stackengine}
\usepackage{bbm}
\usepackage{pifont}
\usepackage[nointegrals]{wasysym}
\usepackage{stmaryrd}
\usepackage{xcolor}
\usepackage{xurl}
\usepackage{stfloats}
\usepackage{todonotes}

\DeclareRobustCommand{\perthousand}{%
  \ifmmode
    \text{\textperthousand}%
  \else
    \textperthousand
  \fi}

\AtBeginDocument{%
  \providecommand\BibTeX{{%
    \normalfont B\kern-0.5em{\scshape i\kern-0.25em b}\kern-0.8em\TeX}}}

\theoremstyle{remark}
\newtheorem{example}{\it Example}

\usepackage{epsfig,amsmath,amsfonts,epsfig,multirow,makecell,caption,soul,csquotes,color,wrapfig,subcaption,mathtools,bm,spverbatim,booktabs,tcolorbox,diagbox}
\usepackage[e]{esvect}

\def\tablename{Table}
\def\figurename{Figure}

\usepackage{caption}
\makeatletter
\newif\if@restonecol
\makeatother

\usepackage[boxed, ruled, vlined, linesnumbered]{algorithm2e}
\SetKwRepeat{Do}{do}{while}

\newenvironment{changemargin}[2]{\begin{list}{}{
	\setlength{\topsep}{0pt}\setlength{\leftmargin}{0pt}
	\setlength{\rightmargin}{0pt}
	\setlength{\listparindent}{\parindent}
	\setlength{\itemindent}{\parindent}
	\setlength{\parsep}{0pt plus 1pt}
	\addtolength{\leftmargin}{#1}\addtolength{\rightmargin}{#2}
	}\item}
	{\end{list}}

\usepackage[first=0,last=9]{lcg}
\usepackage{colortbl}
\definecolor{Gray}{gray}{0.8}
\colorlet{Red}{red!10!white}
\colorlet{Blue}{blue!10!white}

\usepackage{hyperref}

\newcommand{\msec}[1]{\S\ref{#1}}
\newcommand{\mref}[1]{\,\ref{#1}}
\newcommand{\meq}[1]{Eq.\,\ref{#1}}
\newcommand{\mcite}[1]{\cite{#1}}

\newcommand{\meg}{\textit{e.g.}\xspace}
\newcommand{\mie}{\textit{i.e.}\xspace}
\newcommand{\mcf}{\textit{cf}.\xspace}

\newcommand{\mct}[1]{({\it #1})}

\newtcolorbox{mtbox}[1]{left=0.25mm, right=0.25mm, top=0.25mm, bottom=0.25mm, sharp corners, colframe=red!50!black, boxrule=0.5pt, title={#1}, fonttitle=\bfseries, coltitle=red!50!black, attach title to upper={\ --\ }}

\usepackage{scalerel}[2016/12/29]

\makeatletter
\providecommand{\leadsfrom}{%
  \mathrel{\mathpalette\reflect@squig\relax}%
}
\newcommand{\reflect@squig}[2]{%
  \reflectbox{$\m@th#1\leadsto$}%
}
\makeatother

\def\eqref#1{equation~\ref{#1}}

\def\1{\bm{1}}

\DeclareMathAlphabet{\mathsfit}{\encodingdefault}{\sfdefault}{m}{sl}
\SetMathAlphabet{\mathsfit}{bold}{\encodingdefault}{\sfdefault}{bx}{n}

\def\gA{{\mathcal{A}}}

\def\gC{{\mathcal{C}}}
\def\gD{{\mathcal{D}}}
\def\gE{{\mathcal{E}}}

\def\gG{{\mathcal{G}}}

\def\gM{{\mathcal{M}}}
\def\gN{{\mathcal{N}}}

\def\gR{{\mathcal{R}}}

\def\gV{{\mathcal{V}}}

\begin{document}

\newcommand{\kg}{{KG}\xspace}
\newcommand{\kgs}{{KGs}\xspace}
\newcommand{\gnns}{{GNNs}\xspace}

\newcommand{\prvkg}{$\gG_\mathrm{prv}$\xspace} 
\newcommand{\pubkg}{$\gG_\mathrm{pub}$\xspace} 

\newcommand{\surkg}{$\gG^*_\mathrm{pub}$\xspace} 
\newcommand{\prvsur}{$\gG^*_\mathrm{prv}$\xspace} 

\newcommand{\qa}{{\sc QA}\xspace}
\newcommand{\cve}{{CVE}\xspace}
\newcommand{\cves}{{CVEs}\xspace}
\newcommand{\rec}{{\sc Rec}\xspace}
\newcommand{\krl}{{KGR}\xspace}
\newcommand{\kgr}{{KGR}\xspace}
\newcommand{\mrr}{{MRR}\xspace}
\newcommand{\hit}{{HIT@$K$}\xspace}
\newcommand{\hito}{{HIT@$1$}\xspace}
\newcommand{\hitf}{{HIT@$5$}\xspace}

\newcommand{\ndcg}{{NDCG@$K$}\xspace}
\newcommand{\ndcgf}{{NDCG@5}\xspace}

\newcommand{\akp}{{\sc ROAR}$_\mathrm{kp}$\xspace}
\newcommand{\aqp}{{\sc ROAR}$_\mathrm{qm}$\xspace}
\newcommand{\aco}{{\sc ROAR}$_\mathrm{co}$\xspace}
\newcommand{\basone}{{\sc BAS}$_\mathrm{I}$\xspace}
\newcommand{\bastwo}{{\sc BAS}$_\mathrm{II}$\xspace}

\newcommand{\system}{{\sc Kgx}\xspace}

\newcommand{\attack}{{\sc Kgx}\xspace}

\newcommand{\NA}{----}
\newcommand{\kgp}{{\sc Kgp}\xspace} 
\newcommand{\lqe}{{\sc Lqe}\xspace} 
\newcommand{\cop}{{\sc CoP}\xspace} 
\newcommand{\ota}{{OTA}\xspace} 

\newcommand{\vect}[1]{\boldsymbol{#1}} 
\newcommand{\sxrightarrow}[2][]{%
  \mathrel{\text{\footnotesize $\xrightarrow[#1]{#2}$}}%
}

\stackMath
\newcommand\xxrightarrow[2][]{\scriptsize \mathrel{%
  \setbox2=\hbox{\stackon{\scriptstyle#1}{\scriptstyle#2}}%
  \stackunder[0pt]{%
    \xrightarrow{\makebox[\dimexpr\wd2\relax]{$\scriptstyle#2$}}%
  }{%
   \scriptstyle#1\,%
  }%
}}

\newcommand{\tfont}[1]{\textsf{\small #1}}

\let\oldsqsubset\sqsubset
\renewcommand{\sqsubset}[1][0pt]{%
  \mathrel{\raisebox{#1}{$\oldsqsubset$}}%
}

\newcommand{\zhaohan}[1]{\textcolor{black}{#1}\xspace}
\newcommand{\note}[1]{\colorbox{blush}{#1}\xspace}

\newcommand{\vecv}{\vec{v}}

\title{All Your Knowledge Belongs to Us:\\Stealing Knowledge Graphs via Reasoning APIs}
\author{
{\rm Zhaohan Xi}\\
Binghamton University, State University of New York \\
} 

\maketitle

\begin{abstract}
Knowledge graph reasoning (KGR), which answers complex, logical queries over large knowledge graphs (KGs), represents an important artificial intelligence task with a range of applications. Many KGs require extensive domain expertise and engineering effort to build and are hence considered proprietary within organizations and enterprises. Yet, spurred by their commercial and research potential, there is a growing trend to make KGR systems, (partially) built upon private KGs, publicly available through reasoning APIs.

The inherent tension between maintaining the confidentiality of KGs while ensuring the accessibility to KGR systems motivates our study of KG extraction attacks: the adversary aims to ``steal'' the private segments of the backend KG, leveraging solely black-box access to the KGR API. Specifically, we present \attack, an attack that extracts confidential sub-KGs with high fidelity under limited query budgets. At a high level, \attack progressively and adaptively queries the KGR API and integrates the query responses to reconstruct the private sub-KG. This extraction remains viable even if any query responses related to the private sub-KG are filtered. We validate the efficacy of \attack against both experimental and real-world KGR APIs. Interestingly, we find that typical countermeasures (\meg, injecting noise into query responses) are often ineffective against \attack. Our findings suggest the need for a more principled approach to developing and deploying KGR systems, as well as devising new defenses against KG extraction attacks.
\end{abstract}

\section{Introduction}
\label{sec:intro}

Knowledge graphs (\kgs) are structured representations of human knowledge, describing real-world objects, their properties, and their relations. Recent years are witnessing a significant growth of large-scale \kgs in various domains (\meg, MITRE \mcite{mitre-attack}, UMLS \mcite{umls}, and DrugBank \mcite{drugbank}). One major use of such \kgs is to support knowledge graph reasoning (KGR), which answers complex, logical queries over backend \kgs, enabling a range of applications including information retrieval \mcite{gkg-api}, cyber-threat hunting \mcite{cyscale}, biomedical research \mcite{med-kg-embedding}, and clinical decision support \mcite{qiagen}. 

\begin{figure}[!t]
    \centering
    \epsfig{file = 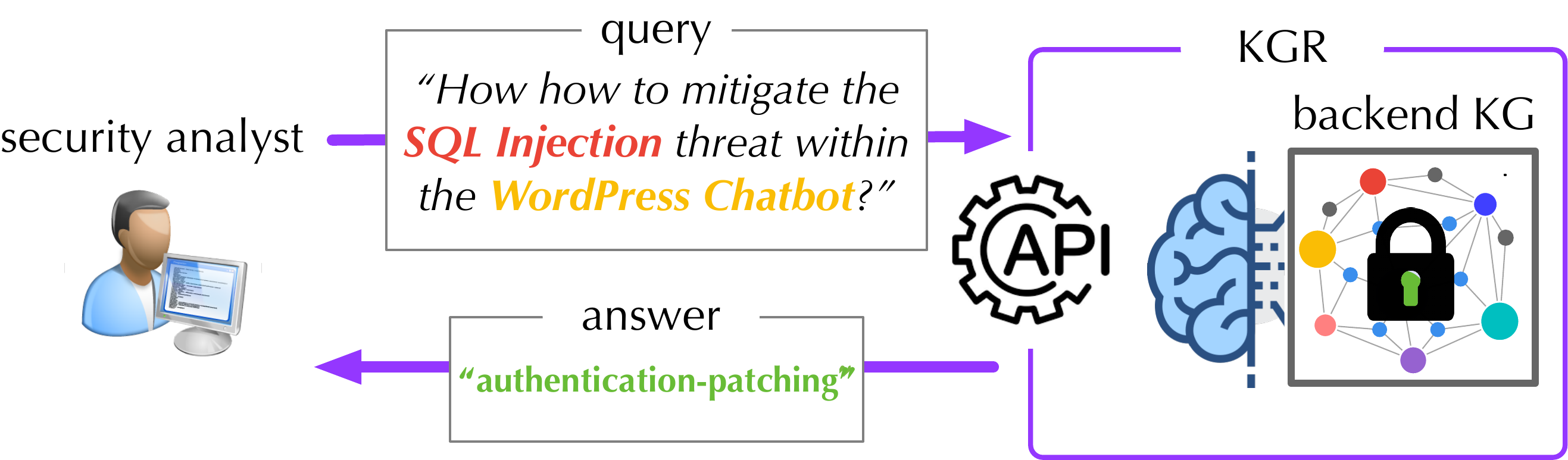, width = 82mm}
    \caption{Illustration of KGR in cyber-threat querying.}
    \label{fig:example}
\end{figure}

Driven by commercial and research potential, there is a growing trend to make KGR  publicly available through reasoning APIs \mcite{gkg-api}. An API encapsulates KGR as a black-box backbone, takes user-provided queries to its KGR for reasoning, and returns answers as output knowledge.
\begin{example}
Figure\mref{fig:example} shows a cyber-threat hunting case: upon observing suspicious network activities, the security analyst may query the \krl system (\meg, LogRhythm \mcite{logrhythm}): ``{\em how to mitigate the \tfont{SQL Injection} threat within the \tfont{WordPress Chatbot}}?'' Processing the query over the backend \kg identifies the most likely answer \tfont{authentication-patching} \mcite{wordpress-chatbot-cve}.
\end{example}

However, providing this public accessibility to KGR tends to conflict with the potentially sensitive nature of backend KGs. Specifically, many KGs require extensive domain expertise and engineering effort to curate and maintain and are thus considered proprietary within organizations and enterprises \mcite{qiagen}. 
Further, as representation learning-based KGR \mcite{gartner-report} often lag behind their backend KGs in terms of updates, their discrepancy entails significant implications. For instance, MITRE accidentally published (and later retracted) the IP addresses of over a dozen unpatched systems \mcite{cve-accident}, which, once exposed by KGR APIs, may jeopardize these vulnerable systems. 



To address the inherent tension between preserving the confidentiality of \kgs and ensuring accessibility, cloud-based platforms, such as Google \mcite{setup-gkg-api} and Amazon Web Services (AWS) \mcite{aws-kg-api}, empower knowledge providers to regulate information accessibility through APIs. These APIs implement several mechanisms to restrict user access, including: (i) a rate limiting to cosntrain users within specific query budgets \mcite{google-kg-usage-limit}; (ii) the service differentiation to segregate services into complimentary and premium (\mie, pay-to-access) tiers \mcite{gkg-edition}; (iii) firewall policies to monitor querying and obscure certain segments of knowledge for security purposes \mcite{johnson2016guide}.

\begin{example}
When a security analyst submits an inquiry to the \kgr system ``{\em which components within the \tfont{WordPress Chatbot} are susceptible to \tfont{SQL injection}}?'' the KGR system may encounter limitations in providing informative responses. This limitation arises from the presence of a firewall mechanism designed to safeguard against the disclosure of unresolved security vulnerabilities \mcite{wordfence-report}. Consequently, the KGR system might be unable to yield accurate information regarding specific vulnerabilities, particularly those related to \tfont{SQL injection}.
\end{example}

\zhaohan{Nevertheless, the delicate balance between upholding the \kg confidentiality and facilitating access control implies the potential information ``leakage'' through APIs. Given well-crafted queries, adversaries could unveil the confidential knowledge from APIs, consequently undermining the integrity of knowledge management. This, in turn, motivates us to explore strategies to extract confidential knowledge through restricted API accessibility.}


In this paper, we study {\em KG extraction attacks}, which explore queries to trigger the information leakage of API, reveal what confidential knowledge potentially present in backend KGs, and ultimately reconstruct (``steal'') confidential sub-KGs via black-box access to KGR APIs. Specifically, we aim to answer the following key questions:

\vspace{1pt}
\noindent RQ$_1$ -- {\em Are KGR APIs susceptible to KG extraction attack?}

\vspace{1pt} 
\noindent
RQ$_2$ -- {\em How effective are the attacks under various practical
settings (\meg, ``open-world'' KGR APIs)?}

\vspace{1pt}
\noindent
RQ$_3$ -- {\em What are the potential countermeasures?} 

However, despite the plethora of work on extracting training data \mcite{model-inversion,zhang2022inference, salem2020updates} or model functionalities \mcite{tramer2016stealing, shen2022model,wang2018stealing} via machine-learning-as-a-service (MLaaS) APIs, KG extraction entails unique challenges: \mct{i} compared with general data (\meg, images), KGs contain much richer relational structures that need to be reconstructed; \mct{ii} \kgr differs significantly from predictive tasks in terms of query processing (details in \msec{sec:background}), precluding distillation-based extraction \mcite{shen2022model}; \mct{iii} unlike predictive tasks, KGR may filter any query answers pertaining to the private sub-KG (details in \msec{sec:threatmodel}).

\vspace{2pt}
{\bf Our work --} This work represents the design, implementation, and evaluation of \attack, the first KG extraction attack against KGR APIs. \attack focuses on uncovering the relational structures of the private sub-KG. At a high level, it progressively and adaptively generates reasoning queries that are likely to intersect with the private sub-KG; it then infers the intersections based on the query answers, even if any answers relevant to the confidential sub-KG are excluded; moreover, it consolidates the intersections of different queries to reconstruct the private sub-KG.  

We validate the practicality of \attack using a range of benchmark KGs, models, and reasoning methods, leading to
the following interesting findings.

\vspace{1pt}
RA$_1$ -- We demonstrate that KGR APIs are highly vulnerable to KG extraction attacks even if any query responses related to private sub-KGs are filtered. For instance, compared with the ground truth, the sub-KG reconstructed by \attack attains over 94\% precision and 90\% recall.

\vspace{1pt}
RA$_2$ -- We further evaluate \attack against KGR APIs ``in the wild'' (\meg, Google KG \mcite{gkg-api}). We show that, despite the enormous size (1.37 million entities and 9.65 million edges) and the open-world nature of Google KG, \attack attains performance comparable with the closed-world setting.  

\vspace{1pt}
RA$_3$ -- Lastly, we discuss potential countermeasures and
their challenges. Although it is straightforward to
conceive high-level mitigation strategies, it is non-trivial to concretely implement them. For instance, injecting random noises into query answers amounts to balancing the delicate trade-off between KGR performance and attack robustness.
Even worse, \attack may neutralize the noises by aggregating related queries, thereby diminishing the effectiveness of such defenses.

\vspace{2pt}
{\bf  Contributions --} To the best of our knowledge, this work represents the first systematic study on the privacy vulnerabilities of KGR APIs to KG extraction attacks. Our contributions are summarized as follows.

\vspace{1pt}
We present \attack, the first KG extraction attack, which highlights the following features: \mct{i} it assumes only black-box query access to the KGR API, \mct{ii} it extracts private sub-KGs that highly align with the ground truth, and \mct{iii} it succeeds under limited query budget.

\vspace{1pt}
We empirically demonstrate the efficacy of \attack against both experimental and real-world KGR APIs across different types of KGs, models, and reasoning methods. The evaluation characterizes the inherent vulnerabilities of KGR APIs to extraction attacks.

\vspace{1pt}
We further discuss potential mitigation and challenges, which sheds light on improving the current practice of operating KGR, pointing to several research directions.

\section{Preliminaries}
\label{sec:background}

This section introduces fundamental concepts and assumptions. The important notations are summarized in Table\mref{tab:notation}.

\begin{table}[!ht]{
\centering
\footnotesize
\renewcommand{\arraystretch}{1.1}
\begin{tabular}{c | l }
Notation & Definition \\
\hline
$\gG$ & backend \kg of the target \kgr API \\
$\gN, \gE, \gR$ & entity, edge, and relation sets of $\gG$\\
\pubkg, \prvkg & public, private sub-KGs of $\gG$: $\gG = \gG_\mathrm{pub}\cup \gG_\mathrm{prv}$\\
$q$, $\llbracket q \rrbracket$ & query, answer\\
$\gG_\mathrm{pub}^*$, $\gG_\mathrm{prv}^*$ & surrogate \kg, extracted sub-KG by the adversary \\
$\gM$ & a matched graph between \prvsur and \prvkg \\
$v_a$, $v_?$ & anchor, answer entities\\
$q(v_a, v_?)$ & relational path from $v_a$ to $v_?$\\
$n_\mathrm{query}$ & query budget\\

\hline
\end{tabular}
\caption{Notations and definitions. \label{tab:notation}}}
\end{table}

\subsection{\kgr} 
\label{ssec:kgr}

{\bf Knowledge graph (\kg)} $\gG = (\gN, \gE)$ consists of node set $\gN$ and edge set $\gE$. Each node $v \in \gN$ represents an entity and each edge $v \sxrightarrow[]{r} v' \in \gE$ indicates that there exists relation $r \in \gR$ (where $\gR$ is a finite set of relation types) from $v$ to $v'$. In other words, $\gG$ comprises a set of facts $\{\langle v, r, v' \rangle \}$ with $v, v' \in  \gN$ and $v \sxrightarrow[]{r} v' \in  \gE$. Additionally, each entity $v$ belongs to a category $c \in \gC$ (where $\gC$ is a finite set of categories).


\vspace{2pt}
{\bf Queries.} A variety of reasoning tasks can be performed over {\kgs} \mcite{lego-icml21,  logic-query-embedding, inductive-reason-icml20}. In this paper, we focus on {\em first-order conjunctive} queries, which ask for entities that satisfy constraints defined by first-order existential ($\exists$) and conjunctive ($\wedge$) logic \mcite{query2box, complex-qa-iclr21, beta-embedding}. Formally, let $\gA_q$ be a set of known entities (anchors), $\gE_q$ be a set of known relations, $\gV_q$ be a set of intermediate, unknown entities (variables), and $v_?$ be the entity of interest. A first-order conjunctive  query 
$q \triangleq (v_?, \gA_q, \gV_q, \gE_q)$ is defined as:
\begin{equation}
\begin{split}
    & \llbracket q \rrbracket = v_? \,.\, \exists \gV_q : \wedge_{v \sxrightarrow[]{r} v' \in \gE_q}  v \sxrightarrow[]{r} v'  \\
    & \text{s.t.} \;\, v \sxrightarrow[]{r} v' = \left\{
    \begin{array}{l}
    v \in \gA_q, v' \in \gV_q \cup \{v_?\}, r\in \gR  \\
     v, v' \in \gV_q \cup \{v_?\}, r\in \gR
    \end{array}
    \right.
\end{split}
\end{equation}
Here, $\llbracket q \rrbracket$ denotes the query answer; the constraints specify that there exist variables $\gV_q$ and entity of interest $v_?$ in the \kg such that the relations between $\gA_q$, $\gV_q$, and $v_?$ satisfy the relations specified in $\gE_q$. 

\begin{example}
\label{exp:query}
The query of ``{\em how to mitigate the \tfont{SQL Injection} threat within \tfont{WordPress Chatbot}?}'' translates into: 
\begin{equation}
\label{eq:sample}
\begin{split}
    q = & (v_?, \gA_q = \{\textsf{\small WordPress Chatbot}, \textsf{\small SQL Injection}\}, \\
    & \gV_q = \{v_\text{threat}\}, \\
    & \gE_q =  \{\textsf{\small WordPress Chatbot} \sxrightarrow[]{\text{target-by}} v_\text{threat}, \\ & \textsf{\small SQL Injection} \sxrightarrow[]{\text{launch-by}} v_\text{threat}, 
    v_\text{threat} \sxrightarrow[]{\text{mitigate-by}} v_? \})
\end{split}
\end{equation}

\end{example}

\vspace{2pt}
{\bf KGR} is essentially a graph-matching problem that matches the entities and relations in queries with that in \kgs. Recently, knowledge representation learning is emerging as a state-of-the-art approach for \krl. It projects \kg $\gG$ and query $q$ to a latent space, such that entities in $\gG$ that answer $q$ are embedded close to $q$. Answering an arbitrary query $q$ is thus reduced to finding entities with embeddings most similar to $q$, thereby implicitly imputing missing relations \mcite{missing-relation} and scaling up to large {\kgs} \mcite{yago}. 

Typically, representation-based \krl comprises two key components: embedding function $\psi$ projects each entity in $\gG$ to its latent embedding; transformation function $\phi$ computes query $q$'s embedding $\psi_q$ based on the embeddings of individual entities in $q$ and their relations. Both functions are typically implemented as trainable neural networks \mcite{logic-query-embedding}.

To process query $q$, one starts from its anchors $\gA_q$ and iteratively applies the transformation function until reaching the entity of interest $v_?$ with the results as $q$'s embedding $\psi_q$. The entities in $\gG$ with the most similar embeddings to $\psi_q$ are then identified as the query answer $\llbracket q \rrbracket$ \mcite{bilinear-embedding}. 

The training of \krl samples a collection of query-answer pairs from the backend \kg $\gG$ as the training set $\gD$ and optimizes $\phi$ and $\psi$ with respect to $\gD$ in a supervised manner. We defer the details of training and reasoning to \msec{sec:kgr_appendix}.

\subsection{Reasoning APIs}

KGR providers often offer black-box APIs for users to issue reasoning queries (\meg, Example\mref{exp:query}). We characterize the KGR API from three aspects:

\vspace{1pt}
{\bf Answer ranking --} As representation learning-based KGR is a probabilistic approximation of exact KG matching, its answer $\llbracket q \rrbracket$ to query $q$ is often in the form of a ranked list of entities. Entities higher up on the list indicate a higher probability of belonging to $q$'s true answer.   

\vspace{1pt}
{\bf Confidence score --} Additionally, the KGR API may also provide confidence scores (\meg, hybrid matching confidence in Google KG \mcite{gkg}) for entities in $\llbracket q \rrbracket$ to measure their likelihood of answering $q$. We assess the impact of confidence scores on KG extraction attacks in \msec{ssec:cm}. 

\begin{figure}[!ht]
    \centering
    \epsfig{file = 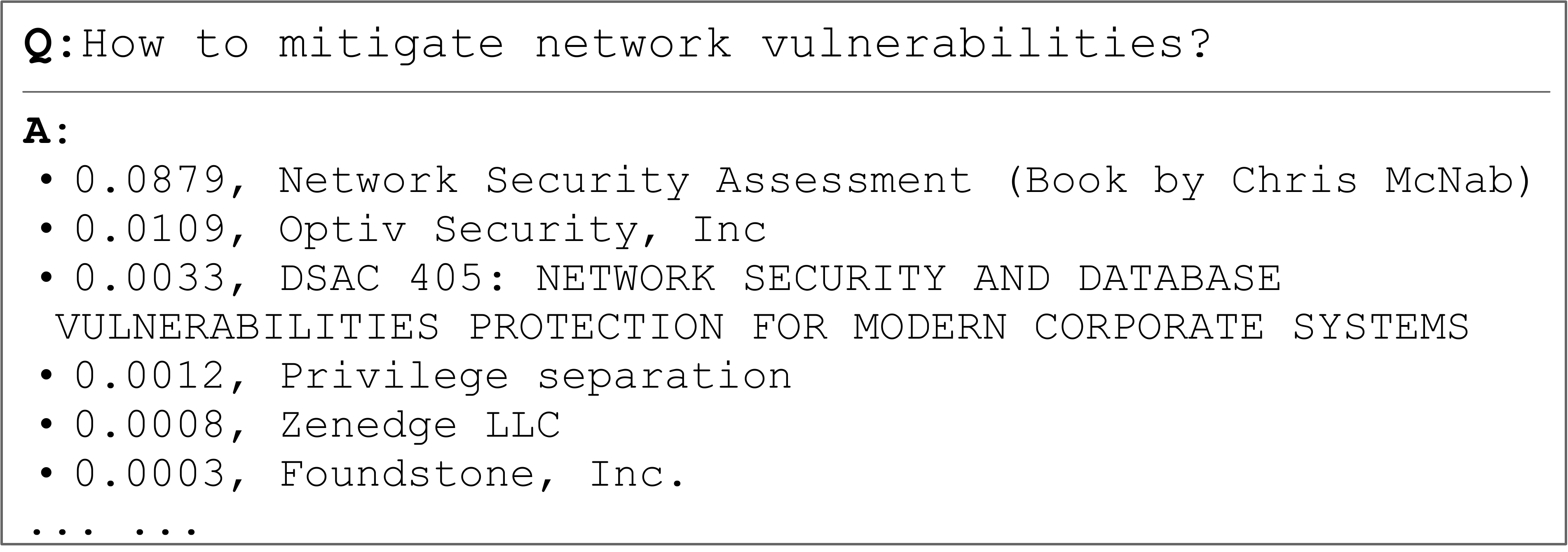, width =85mm}
    \caption{An example of query results using Google KG's API.}
    \label{fig:api}
\end{figure}

\vspace{1pt}
\textbf{Query budget --} The KGR API may also charge the user on a per-query basis. Below, we assume there is a capped total number of queries the adversary is able to issue (\meg, 100,000 daily free queries for Google KG \mcite{google-kg-usage-limit}). Figure\mref{fig:api} shows an example by querying ``{\em how to mitigate network vulnerabilities?}'' to Google KG's API. The returned answer list is ranked by likelihood scores.

\subsection{Private/Public sub-KGs}
\label{ssec:knoweldge_type}

To model the confidentiality of KG $\gG$, without loss of generality, we assume that $\gG$ comprises public sub-KG \pubkg and private sub-KG \prvkg: $\gG = \gG_\mathrm{pub} \cup \gG_\mathrm{prv}$. 

\vspace{1pt}
{\bf Public sub-KG} \pubkg refers to the segment of $\gG$ containing non-sensitive information that is typically accessible without any limitations. For instance, in the context of cyber KGs, \pubkg could include entities from the \tfont{CVE} category, but not from the \tfont{affected-system} category \mcite{cve-accident}; in the case of biomedical KGs, \pubkg may include relations between \tfont{Drug} and \tfont{Disease}, but exclude the relations between \tfont{Drug} and \tfont{Gene} due to privacy concerns \mcite{bonomi2020privacy}.

\vspace{1pt}
{\bf Private sub-KG} \prvkg refers to the fraction of $\gG$ which contains sensitive information and can only be accessed with specific permissions. For example, providers of KGR services may demand licenses to query entities/relations from certain categories \mcite{qiagen}; they may also implement service differentiation \mcite{cyc}, setting a higher price for access to a more sensitive sub-KG \prvkg. Additionally, as representation learning-based KGR models often lag behind their underlying KGs in terms of updates \mcite{cve-accident}, a sub-KG that has been removed during an update can also be deemed as \prvkg.

\begin{figure}[!ht]
    \centering
    \epsfig{file = 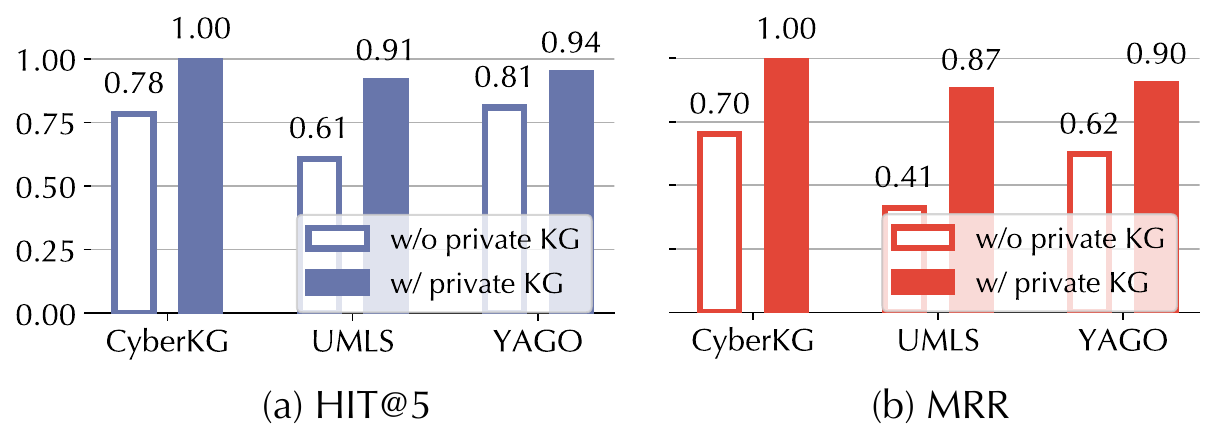, width =82mm}
    \caption{\kgr performance on different \kgs with and without \prvkg, measured by metrics (a) \hitf and (b) \mrr.}
    \label{fig:prv}
\end{figure}

\vspace{1pt}
{\bf Confidentiality of private sub-KG --} An apparent approach to safeguarding the confidentiality of \prvkg is to entirely omit \prvkg during the training of KGR. However, this strategy might not be practical in real-world applications.

First, incorporating \prvkg during the training of KGR may be essential for its performance. Intuitively, the additional entities in \prvkg and their relations within \prvkg and with \pubkg contribute to the effective training of the embedding and transformation functions. Following the setting of \msec{sec:expt}, we train two separate KGR models with and without \prvkg, respectively, and assess their performance in answering queries related to entities in \pubkg. Figure\mref{fig:prv} illustrates the performance disparity caused by \prvkg, measured by \hitf and \mrr
\footnote{MRR is the average reciprocal ranks of ground-truth answers, which measures the global
ranking quality of KGR. HIT@$K$ is the ratio of top-$K$
results that contain ground-truth answers.} 
on three distinct KGs. It is worth noting that the inclusion of \prvkg in KGR training significantly enhances its performance in answering queries solely related to \pubkg (\meg, over 50\% increase in \mrr on UMLS). Furthermore, in the case that the KGR model lags behind the backend KG, eliminating the influence of the deleted sub-KG \prvkg equates to re-training the KGR model, which is often costly and executed only periodically.

Hence, a more pragmatic approach is to filter out any query responses related to \prvkg. Specifically, in the case of first-order conjunctive queries, entities in the query response $\llbracket q \rrbracket$ that belong to \prvkg are excluded from $\llbracket q \rrbracket$.

\section{Threat Model}
\label{sec:threatmodel}

We describe the threat model of KG extraction attacks according to the
adversary’s objectives, knowledge, and capability.

\subsection{Adversary's objectives}
\label{ssec:objective}

The adversary's objective is to reconstruct (``steal'') the private sub-KG \prvkg through querying the KGR API. Formally, let \prvsur denote the private sub-KG reconstructed by the adversary. The attack is successful if \prvsur is highly similar to the ground-truth \prvkg: \prvsur$\approx$\,\prvkg. 

However, as the API excludes any entities in \prvkg from the query answers, it is infeasible to reconstruct the concrete entities in \prvkg (\meg, entity names). Rather, we focus on recovering the relational structures of \prvkg, including the relations between different nodes as well as their categories. 

\begin{figure}[!ht]
    \centering
    \epsfig{file = 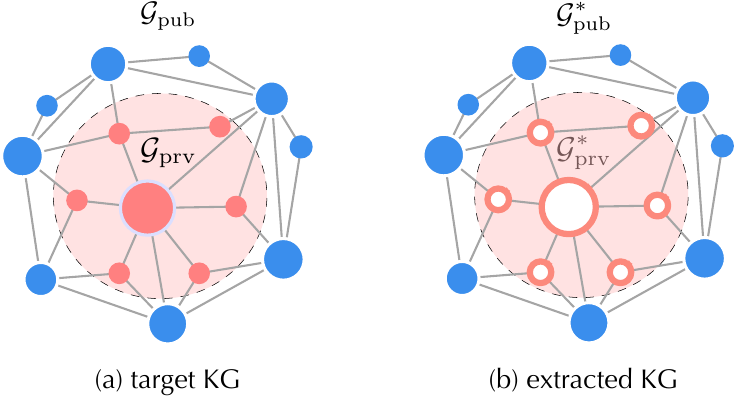, width =72mm}
    \caption{Sample \kg with private sub-KG: (a) target \kg $\gG$ (private sub-KG \prvkg in red); (b) extracted sub-KG \prvsur (in red).}
    \label{fig:prvkg_example}
\end{figure} 



Figure\mref{fig:prvkg_example} illustrates the \kg we expect to extract. Note that the entities in \prvkg are excluded from the query answers, hence unobservable by the adversary; thus, the adversary considers variables (hollow red circles in Figure\mref{fig:prvkg_example}-(b)) as nodes in \prvsur and aims to extract their surrounding relations, such that \prvsur and \prvkg share identical relational structures.


Formally, we denote $\gM$ as a matched graph between \prvsur and \prvkg. To find the maximum match with at most facts, we initialize the matching $\gM$ as entities common in $\gG^*_\mathrm{pub}$ and \pubkg, and gradually grow $\gM$ by following the relations in \prvsur and \prvkg: a new pair of entities ($v^* \in$ \prvsur, $v \in$  \prvkg) are added to $\gM$ only if their relations with existing entities in $\gM$ are matched. Note that we only match each fact at most once.

\begin{example}
    Assuming \prvkg contains a fact \tfont{Chatbot}$\sxrightarrow[]{\text{target-by}}$\tfont{Unauthenticated Attack}, where \tfont{Chatbot} is in \pubkg and \tfont{Unauthenticated Attack} is in \prvkg. An extracted fact is \tfont{Chatbot}$\sxrightarrow[]{\text{target-by}}v_\text{threat}$ with variable $v_\text{threat}$ in \prvsur. We calculate $\gM$ by first finding their common entities in \pubkg,  \mie, \tfont{Chatbot}, then following the common relation \tfont{target-by} to match next-hop entities respectively in \prvsur and \prvkg, \mie, we match $v_\text{threat}$ and \tfont{Unauthenticated Attack}. Finally, we denote the extracted fact \tfont{Chatbot}$\sxrightarrow[]{\text{target-by}} v_\text{threat}$ is matched with \tfont{Chatbot}$\sxrightarrow[]{\text{target-by}}$ \tfont{Unauthenticated Attack}. Furthermore, we can start with $v_\text{threat}$ and \tfont{Unauthenticated Attack} to match more facts between \prvsur and \prvkg.
\end{example}

\zhaohan{It is important to note that the matching described above is primarily employed for assessing the similarity between \prvkg and the extracted \prvsur as discussed in Section \msec{sec:expt}. In practical extraction scenarios, the adversary must rely on auxiliary sources of knowledge, such as cyber-threat reports or clinical literature, in order to identify specific entities within the \prvkg.}



\subsection{\zhaohan{Subsequent impacts}}

\zhaohan{Note that the relational structures of \prvkg capture the essential semantics of \prvkg, which support knowledge representation learning to generate the semantic embeddings of entities \mcite{query2box, kg-reasoning-review, beta-embedding, logic-query-embedding}. Thus, uncovering the relational structures of \prvkg raises several subsequent impacts as adversary's next-step objectives, specifically:}

\zhaohan{\textbf{1. Commercial crisis --} With equivalent \kg structures, an adversarial is capable of developing an alternative \kgr service that can match the performance of the authentic service. This scenario introduces a challenge to the pay-to-access privilege and poses a risk to IP protection.}

\zhaohan{\textbf{2. Subsequent attacks --} Exposing the structural information within the \kg is sufficient to facilitate follow-up attacks, including the creation of deceptive or ``poisoning'' knowledge \mcite{kg-attack}, even in cases where the specific terminology of entities remains undisclosed.}

\zhaohan{\textbf{3. Entity resolution --} With access to supplementary domain knowledge, an adversarial can potentially establish connections between nodes within \prvsur and specific, tangible entities. For example, it becomes feasible to identify a concrete malware entity within \prvsur by analyzing its relationships in auxiliary threat reports. This discovery may subsequently incentivize the adversary to exploit potentially vulnerable software by leveraging the extracted \prvsur structures that connect to the resolved ``malware'' entity.}

\zhaohan{Despite the above implications, this work mainly focuses on the step of uncovering the relational structures of \prvkg and considers its subsequent impacts as ongoing research.}

\subsection{Adversary's knowledge}

We now consider the adversary's prior knowledge regarding the target \kg $\gG$ and the KGR model.

\zhaohan{\textbf{Public knowledge --} Recall that $\gG$ comprises public sub-KG \pubkg and private sub-KG \prvkg, wherein obtaining an adequate part of \pubkg is feasible in two ways. First, real-world \kgs are often crowdsourced from public, free knowledge bases, such as  MITRE \mcite{mitre-attack} for building ThreatKG and UMLS \mcite{umls} for building clinical KG. It is thus easy to construct a large-scale KG with adequate entities. Second, the adversary is able to query the unconstrained and free parts through the KGR API, thereby disclosing a part of the public sub-KG. By combining those two strategies, the adversary is able to collect an adequate part of a public sub-KG to start the attack with.}

\zhaohan{For the convenience of discussion, we temporarily assume the adversary has partial access to \pubkg. Note that this assumption is not necessary. We discuss how to bootstrap the attack without access to \pubkg in \msec{ssec:expt_factor}. Further, we assume the adversary has zero knowledge about \prvkg.}

\zhaohan{\textbf{Private entity category --} Real-world \kgs typically organize their entities as a set of entity groups (e.g., threats, mitigations, or products), while the relation types are often formed between different groups (e.g., \tfont{mitigate-by} relations between threats and mitigations). Thus, the adversary can sample queries to reveal the entity categories in \prvkg. For instance, a query using evidence of cyber threats to ask for mitigations can extract concrete answers, but directly asking for ``what cyber threats are related to given evidence'' may lead to low-confident answers. In this way, we reveal that the ``threat'' category contains partial \prvkg.}

\textbf{\kgr\xspace--} To make the attack more practical, we assume the adversary has no knowledge regarding the KGR system including its underlying models (\meg, GNN \mcite{qa-gnn}), training and reasoning methods (\meg, Query2Box \mcite{query2box} versus GQE \mcite{logic-query-embedding}), and configurations (\meg, embedding dimensionality).

\subsection{Adversary's capability}

\textbf{API access --} we assume the adversary solely relies on black-box query access to the KGR API to perform the attack. Any entities in \prvkg are purged from the query answers. Further, the adversary is constrained by a query budget $n_\mathrm{query}$ that limits the total number of queries (details in \msec{ssec:expt_factor}).

\section{KGX Attack}
\label{sec:attack}

\begin{figure*}[!ht]
    \centering
    \epsfig{file = 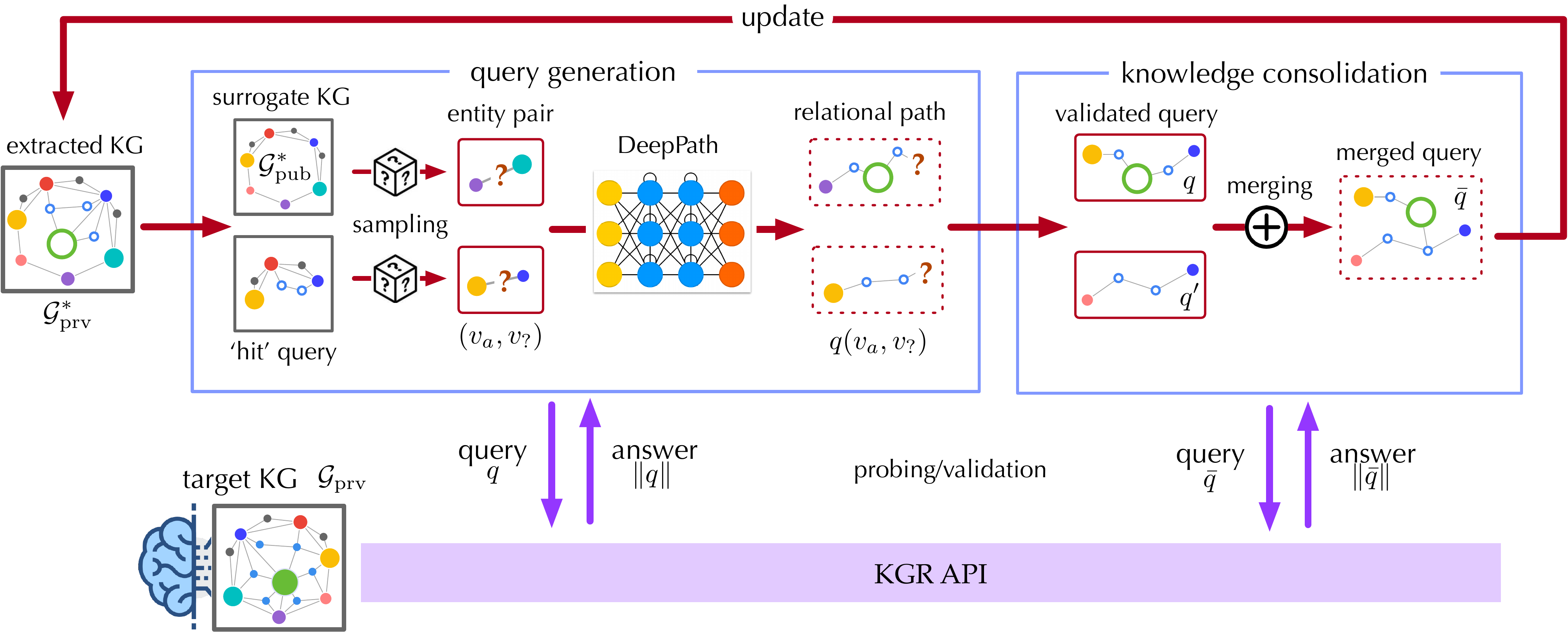, width =176mm}
    \caption{Overview of \attack attack.}
    \label{fig:attack_overview}
\end{figure*}

We now present \attack, the first KG extraction attack against KGR APIs. 

\subsection{Attack overview}

As illustrated in Figure\mref{fig:attack_overview}, \attack reconstructs the private sub-KG through progressively and iteratively querying the KGR API and integrating the query responses. In a nutshell, \attack iterates between two steps: query generation and knowledge consolidation, as described below. 

\vspace{2pt}
{\bf Query generation} adaptively generates queries that potentially intersects with \prvkg. Using such queries to ``probe'' the KGR API may reveal the relational structures of \prvkg. Specifically, one may adopt two strategies to generate query $q$: \mct{i} \underline{\em exploitation} generates $q$ in the vicinity of the already-uncovered segment of \prvkg, which has a high chance of intersecting with \prvkg but may only reveal limited new information; \mct{ii} \underline{\em exploration} generates $q$ away from the already-uncovered segment of \prvkg, which incurs the risk of missing \prvkg but, if intersected, may effectively reveal unknown structures of \prvkg. In \attack, we seek to balance these two strategies to maximize the query efficiency.

\vspace{2pt} 
{\bf Knowledge consolidation} integrates the responses of multiple queries to infer the unknown structures of \prvkg. Given a query $q$ that intersects with \prvkg, even if any entities in \prvkg are excluded from $q$'s answer $\llbracket q \rrbracket$, it remains feasible to infer \mct{i} whether $q$ intersects with \prvkg and \mct{ii} what is the intersected structure. By merging such intersected structures from multiple queries, \attack progressively uncovers the relational structures of \prvkg. Note that the adversary may re-query the KGR API to validate this merging operation.

\begin{algorithm}[!t]{
\footnotesize
\KwIn{
    \surkg -- initial surrogate KG; $n_\mathrm{query}$ -- query budget
}
\KwOut{
    \prvsur -- extracted private sub-KG;
}

\prvsur$ \leftarrow \emptyset$\;
\tcp{bootstrapping}
expand \surkg by querying API\;
update $n_\mathrm{query}$\;
  \tcp{KG extraction}
\While{$n_\mathrm{query} > 0$}{
    \tcp{query generation}
    $(v_a, v_?) \sim $  \surkg (exploration) or ``hit'' queries (exploitation)\;
    \ForEach{entity pair $(v_a, v_?)$}{
        path $q(v_a, v_?) \leftarrow \textsc{DeepPath}(v_a, v_?)$\;
        validate $q(v_a, v_?)$ by querying API\; 
    }
    \tcp{knowledge consolidation}
    \ForEach{pair of validated queries $(q, q')$}{
         $\bar{q} \leftarrow$ merging $q$ and $q'$\;
         validate $\bar{q}$ by querying API\; 
         \lIf{validated}{update \prvsur with $\bar{q}$}
    }
    update $n_\mathrm{query}$\;
}
\Return \prvsur\;
\caption{\system Attack. \label{alg:attack}}}
\end{algorithm}



We elaborate on query generation and knowledge consolidation in \msec{ssec:query_gen} and \msec{sec:consolid}, respectively. Further, we lift the assumption that the adversary has partial access to public sub-KG \pubkg and show how to bootstrap \attack without knowledge about \pubkg in \msec{ssec:bootstrap}. 
Putting everything together, Algorithm\mref{alg:attack} sketches the overall flow of \attack.

\subsection{Query generation}
\label{ssec:query_gen}

Recall that first-order conjunctive queries ask for entities that satisfy existential and conjunctive constraints. While many types of queries are possible, we focus on generating path queries that specify a sequence of conjunctive constraints from an anchor entity $v_a$ to an answer entity $v_?$:
\begin{equation}
\hspace{-3pt}
\llbracket q \rrbracket = v_? \,.\, \exists \{ v_i\}_{i=1}^h : v_a \sxrightarrow[]{r_0} v_1 \wedge v_1 \sxrightarrow[]{r_1} v_2 \wedge \ldots \wedge v_{h} \sxrightarrow[]{r_h} v_? 
\end{equation}
For instance, $\textsf{\small BusyBox} \xxrightarrow[]{\text{target-by}} v_\text{malware}  \xxrightarrow[]{\text{mitigate-by}} v_?$ is a path query asking for the mitigation for malware targeting \tfont{BusyBox}. We choose path queries due to their simplicity and the flexibility of merging multiple queries (details in \msec{sec:consolid}). Note that we expect path queries to go through \prvkg, \mie, the anchor and answer are in \prvkg, while intermediate variables are in \prvkg. Therefore, we generate path queries with at least two-hop relations to guarantee such nature.


To generate a path query, one needs to specify \mct{i} the anchor entity $v_a$ and the (expected) answer entity $v_?$ and \mct{ii} the relational path from $v_a$ to $v_?$ (\mie, the sequence of relations). We first answer the question of how to generate the relational path from $v_a$ to $v_?$ assuming $v_a$ and $v_?$ are given.

\vspace{2pt}
{\bf Generating relational paths.} Given entity pair $(v_a, v_?)$, we employ a model to suggest a likely relational path from $v_a$ to $v_?$. Specifically, we extend {\sc DeepPath} \mcite{deeppath}, a reinforcement learning framework for learning multi-hop relational paths in KGs. Intuitively, {\sc DeepPath} trains a policy network that, given an entity pair, iteratively extends the most likely path between them. However, it is infeasible to directly apply {\sc DeepPath}, as it assumes access to KG $\gG$ and iteratively updates the path by traversing $\gG$, which is inaccessible in \attack. To address this limitation, we train a set of relation-specific projection operators $\{\phi_r\}_r$, in which $\phi_r(\vec{v})$ computes the embeddings of entities with relation $r$ to $v$ ($\vec{v}$ denotes the embedding of entity $v$). Similar to  \mcite{query2box}, the projection operators are implemented as trainable neural networks. 

Under this setting, the state vector is defined as:
\begin{equation}
s = ( \vec{v}, \vec{v}_? - \vec{v} )
\end{equation}
where $\vec{v}$ is the embedding of the current entity $v$ (which is initialized as the anchor entity $v_a$). The action is defined as extending a relation $r \in \gR$ from $v$, with the embedding of the current entity updated as $\phi_r (\vec{v})$ and the state updated as:
\begin{equation}
s' = (\phi_r(\vec{v}), \vec{v}_? - \phi_r(\vec{v}))
\end{equation}
The process iterates until $v_?$ is reached (\mie, $\|\vec{v}_? - \phi_r(\vec{v})\| \leq \epsilon$) or the search fails (\mie, the maximum number of hops $n_\mathrm{hop}$ is reached). If successful, a relational path $q(v_a, v_?) = v_a \sxrightarrow[]{r_0} v_1 \ldots  v_h \sxrightarrow[]{r_{h}} v_?$ is suggested.\footnote{Below we abuse the notation of $q$ for both a query and its relational path.} Given the probabilistic nature of {\sc DeepPath}, we may generate multiple distinctive paths to ensure the diversity of the generated queries. 

Also, note that $q(v_a, v_?)$ may partially overlap with public sub-KG \pubkg. To optimize the distinguishing power of the generated query, we trim the relations trailing $v_a$ (or preceding $v_?$) that intersect with \pubkg and update the anchor (or answer) entity accordingly. For instance, if $q(v_a, v_?) = v_a \sxrightarrow[]{r_0} v_1 \sxrightarrow[]{r_1} v_2 \sxrightarrow[]{r_2} v_?$ and the first hop $v_a \sxrightarrow[]{r_0} v_1$ overlaps with \pubkg, we update the anchor entity $v_a$ as $v_1$ and trim the relational path as $q(v_a, v_?) = v_a \sxrightarrow[]{r_1} v_2 \sxrightarrow[]{r_2} v_?$.

\vspace{2pt}
{\bf Generating entity pairs.} Next, we answer the question of how to generate the entity pair $(v_a, v_?)$. We consider two potential strategies. 

\vspace{1pt}
\underline{\em Exploration} generates $(v_a, v_?)$ by randomly sampling a pair of entities from \pubkg. Due to the randomness, it is likely that \mct{i} there may not exist a relational path $q(v_a, v_?)$ and \mct{ii} even if $q(v_a, v_?)$ exists, it may not overlap with \prvkg. However, given that $q(v_a, v_?)$ exists and intersects with \prvkg, it tends to reveal previously unknown structures of \prvkg.

\vspace{1pt}
\underline{\em Exploitation} generates $(v_a, v_?)$ by modifying a ``hit'' query $q$ (\mie, $q$'s relational path exists and overlaps with \prvkg). Specifically, let $v_a^q$ ($v_?^q$) be $q$'s anchor (answer) entity. We search \pubkg for the nearest neighbor of $v_a^q$ ($v_?^q$) in the embedding space as $v_a$ ($v_?$). Further, we ensure that $v_a$ and $v_a^q$ ($v_?$ and $v_?^q$) share the same category. As $q$ is a hit query, the path from $v_a$ to $v_?$ tends to also exist and  intersect with \prvkg. However, due to its proximity to $q$, this path may only reveal limited new information about \prvkg. 

\vspace{1pt}
To leverage both strategies, \attack interleaves between exploration and exploitation queries.

\subsection{Knowledge consolidation}
\label{sec:consolid}

We use the queries generated in the previous step to probe the KGR API. Below we first describe how to utilize a single query to infer one path in \prvkg and then show how to integrate the paths uncovered by multiple queries to form more complex structures.

\vspace{2pt}
{\bf Utilizing individual queries.} Recall that a path query specifies a sequence of relations $q(v_a, v_?) = v_a \sxrightarrow[]{r_0} v_1 \ldots  v_h \sxrightarrow[]{r_{h}} v_?$, while $q$'s answer $\llbracket q \rrbracket$ returned by the API is a ranked list of entities with confidence scores. We determine the existence of $q(v_a, v_?)$ by checking \mct{i} whether $v_?$ appears in $\llbracket q \rrbracket$ and \mct{ii} whether its confidence score exceeds a threshold $\lambda$.

We set $\lambda$ empirically to differentiate between true and false answers. Specifically, we sample from \pubkg a set of path queries and their answers as the positive cases $\gD^+$, and then perturb the relations or replace the answers (with irrelevant ones) in $\gD^+$ to form the negative cases $\gD^-$. We set $\lambda$ to maximize the difference between the answers in $\gD^+$ and $\gD^-$. More details are deferred to \msec{ssec:score}.

Another way to validate the path between $v_a$ and $v_?$ (without confidence scores) is to query the path reversely. Specifically, we switch the roles of $v_a$ and $v_?$ ($v_a$ as the answer and $v_?$ as the anchor) and query the path $q(v_?, v_a)$. If $v_a$ also appears in $\llbracket q \rrbracket$, we admit the existence of the path from $v_a$ to $v_?$.   

Once we determine that $v_?$ is a true answer for $q$, we consider that the path $v_1 \sxrightarrow[]{r_1} v_2 \ldots  v_{h-1} \sxrightarrow[]{r_{h-1}} v_h$  exists in \prvkg.\footnote{In general, $v_1 \sxrightarrow[]{r_1} v_2 \ldots  v_{h-1} \sxrightarrow[]{r_{h-1}} v_h$ may intersect with \prvkg at disjoint segments, which however is unlikely for a small $h$ (\meg, $h \leq 3$ in our  setting).}

\vspace{2pt}
{\bf Integrating multiple queries.} We may further integrate the hidden paths uncovered from multiple queries into conjunctive forms, revealing more complex  structures of \prvkg.

Specifically, we examine queries that share the same anchor $v_a$ (or answer $v_?$) and greedily merge the common relations trailing $v_a$ (or preceding $v_?$) in their paths. For instance, consider queries $q$ and $q'$ with paths $v_a \sxrightarrow[]{r_0r_1r_2} v_?$ and 
$v_a' \sxrightarrow[]{r_0'r_1'r_2'} v_?'$ (intermediate variables are omitted), respectively. As shown in \meq{eq:merge}, if $v_? = v_?'$ and $r_2 = r_2'$, $q$ and $q'$ are merged as \mct{i}; if $v_a = v_a'$ and $r_0 = r_0'$, $q$ and $q'$ are merged as \mct{ii}. 
\begin{equation}
\label{eq:merge}
(i)\;\; \begin{array}{c|}
v_a \xrightarrow[]{r_0r_1}   \\
v_a' \xrightarrow[]{r_0'r_1'} 
\end{array} \xrightarrow[]{r_2} v_?  \qquad (ii) \;\; v_a \xrightarrow[]{r_0}  
\begin{array}{|c}
\xrightarrow[]{r_1r_2} v_?  \\
 \xrightarrow[]{r_1'r_2'} v_?'
\end{array}
\end{equation}

To validate this merge operation, we may query the  API using the merged structure. In the case of \mct{i} in \meq{eq:merge}, the new structure represents a valid first-order conjunctive query (with anchors $\{v_a, v_a' \}$); we can thus directly query the API to check whether $v_?$ appears in the query answer. In the case of \mct{ii} in \meq{eq:merge}, the new structure does not correspond to a first-order conjunctive query. However, we may switch the roles of anchor and answer entities by considering $v_a$ as the answer entity and \{$v_?$, $v_?'$\} as the anchor entities. This makes the new structure a valid first-order conjunctive query, which can be readily checked.

\subsection{Bootstrapping}
\label{ssec:bootstrap}
Thus far we assume the adversary knows the public sub-KG \pubkg to launch \attack. Here, we lift this assumption and show how to bootstrap the attack. 

Instead, we assume the adversary knows a set of public entities in \pubkg. We argue that this assumption is minimal and practical, as large \kgs are often built upon multiple sources, including public sources. For instance, Google KG partially inherits Freebase \mcite{freebase-dump}, a public KG,

Starting with this set of entities as a seed KG \surkg, we may query the API to reconstruct \pubkg. Specifically, for each seed $v$, we use it as the anchor, sample a relation $r \in \gR$, and generate a one-hop query $q(v, v_?) = v \sxrightarrow[]{r} v_?$. 
By querying the API with $q$ and filtering out entities in $\llbracket q \rrbracket$ with confidence scores below a threshold $\lambda$, we add the remaining entities to \surkg (with relation $r$ to $v$). By iteratively executing the above  steps, we expand \surkg to approach \pubkg. Note that it often suffices for \attack to work that \surkg covers a fraction (\meg, 50\%) of \pubkg (details in \msec{sec:expt}).  
\section{Evaluation}
\label{sec:expt}

Next, we conduct an empirical evaluation of \attack to answer the following key questions. 

\vspace{1pt}
\noindent
Q$_1$ -- Is \attack effective in KG extraction?

\vspace{1pt}
\noindent
Q$_2$ -- What factors influence its performance?

\vspace{1pt}
\noindent
Q$_3$ -- What is the best practice to operate \attack?

\vspace{1pt}
\noindent
Q$_4$ -- How does \attack perform in real-world settings?  

\subsection{Experimental setup}
\label{ssec:expt:setting}

We begin by describing the experimental setting.

\vspace{1pt}
{\bf KG --} We consider four representative \kgs from different domains. \mct{i} CyberKG \mcite{kg-attack} integrates CVE \mcite{cve}, MITRE ATT\&CK \mcite{mitre-attack}, and NVD \mcite{nvd}, and supports cyber-threat hunting; \mct{ii} UMLS \mcite{umls} integrates terminologies from clinical resources to facilitate biomedical research; \mct{iii} YAGO \mcite{yago} is a commonsense \kg benchmark collected from online sources; and \mct{iv} Google \kg \mcite{gkg} is the backbone of Google Knowledge Search API \mcite{gkg-api}. As Google \kg is not directly accessible, we detail how to extract a sub-KG for  evaluation in \msec{ssec:case_study}. Table\mref{tab:kg} summarizes the KG statistics. We use CyberKG, UMLS, and YAGO to evaluate \attack in Q$_1$--Q$_3$ and use Google \kg as a real-world case study in Q$_4$.

\begin{table}[!ht]{
\centering
\footnotesize
\setlength{\tabcolsep}{2pt}
\renewcommand{\arraystretch}{1.3}
\begin{tabular}{c|c|c|c|c}
\multirow{2}{*}{\kg} & $|\gN|$ & $|\gR|$ & $|\gE|$ & \multirow{2}{*}{\kgr Model} \\
& (\#entities) & (\#relations) & (\#facts) & \\
\hline
CyberKG & 178k & 23 & 1.56m & Q2B \\
UMLS & 400k & 176 & 4.84m & GQE \\
YAGO & 300k & 78  & 1.05m & BetaE  \\
Google \kg & 1.37m & 780 & 9.65m & \multicolumn{1}{c}{Google Knowledge Panel} \\
\end{tabular}
\caption{Statistics of KGs and corresponding \kgr models.\label{tab:kg}}}
\end{table}

\vspace{1pt}
{\bf \kgr model --} To rule out the bias of specific models, we consider a variety of KGR models. In Q$_1$--Q$_3$, we use three knowledge representation-based models from SNAP,\footnote{SNAP: \url{https://snap.stanford.edu/}} GQE \mcite{logic-query-embedding}, Q2B \mcite{query2box}, and BetaE \mcite{beta-embedding}, with the varying design of reasoning mechanisms, embeddings, and transformation functions. To answer Q$_4$, we use Google's Knowledge Search API (without access to its KGR model). 

\vspace{1pt}
{\bf Private sub-KG --} \prvkg comprises entities (and their relations) from sensitive categories: \tfont{vulnerability} in CyberKG, \tfont{gene} in UMLS, and \tfont{therapy} in YAGO and Google KG as shown in Table\mref{tab:expt_main}. By default, we sample \prvkg as a connected sub-KG with entities from a single category. We consider other settings that \prvkg includes entities from alternative categories (Table\mref{tab:prv_type}) or a set of disconnected small sub-KGs (Table\mref{tab:rand_prv}). As discussed in \msec{sec:background}, \prvkg is used in KGR training, but any entities in \prvkg are excluded from the query answers.

\vspace{1pt}
{\bf API --} In Q$_1$--Q$_3$, we use the logic query interfaces provided by SNAP to simulate the KGR APIs. In Q$_4$, we directly use Google Knowledge Search API \mcite{gkg-api}. We assume the APIs provide top-$k$ ($k = 10$ by default) entities as query answers, associated with their confidence scores. The definitions of confidence scores vary with concrete KGR models. For instance, SNAP models use distance-based metric (\meg, $L_1$ norm) to measure confidence, while Google KG provides matching scores that represent ``a combination of multiple confidence measurements \mcite{gkg-api-score}''. 

\vspace{1pt}
{\bf KGX --} By default, we assume the adversary has access to a surrogate KG $\gG^*_\mathrm{pub}$ that includes 50\% facts of the public sub-KG \pubkg. We consider the setting that the adversary has little knowledge about \pubkg in \msec{ssec:expt_factor}. Further, the total number of queries the adversary is able to issue is limited by a query budget $n_\mathrm{query}$. By default, we set $n_\mathrm{query}$ as 500K for CyberKG and YAGO, and 1,000K for UMLS and Google KG. We allocate half of the budget to expand surrogate KG $\gG^*_\mathrm{pub}$ and use the remaining budget for KG extraction attacks. The other default setting of \attack is summarized in \msec{sec:param_appendix}.

\vspace{1pt}
{\bf Metrics --} To evaluate the effectiveness of \attack, we compare the extracted sub-KG \prvsur with the ground-truth \prvkg. As the query answers exclude entities in \prvkg, it is infeasible to directly compare \prvsur and \prvkg. We thus resort to finding a maximum matching between \prvsur and \prvkg in terms of their relational structures, defined as their maximum common edge subgraphs \mcite{max-subgraph-wiki}. 
As introduced in \msec{ssec:objective}, we initialize the matching $\gM$ as entities common in $\gG^*_\mathrm{pub}$ and \pubkg, and gradually grow $\gM$ by following the relations in \prvsur and \prvkg.

We measure the matching $\gM$ using the following quantities: \mct{i} $n_\mathrm{prv}$ -- the number of entities (facts) in \prvkg, \mct{ii} $n_\mathrm{prv}^*$ -- the number of entities (facts) in \prvsur, \mct{iii} $n_\mathrm{match}$ -- the number of entities (facts) in $\gM$, \mct{iv} $n_\mathrm{match} / n_\mathrm{prv}^*$ -- the precision of $\gM$, and \mct{v} $n_\mathrm{match} / n_\mathrm{prv}$ -- the recall of $\gM$. In addition, we use graph edit distance (GED) \mcite{ged}, which is defined as the minimum cost of transforming one graph to another, to measure the overall similarity of \prvsur and \prvkg. 
 

\begin{table*}[!ht]{
\centering
\footnotesize
\setlength{\tabcolsep}{5pt}
\renewcommand{\arraystretch}{1.4}
\begin{tabular}{c|c|c|c|c|c|c|c|c|c|c|c|c}
 \multirow{2}{*}{\kg} & Entity Category  &\multicolumn{5}{c|}{Entity} & \multicolumn{5}{c|}{Fact} &\multirow{2}{*}{GED}\\
 \cline{3-12}
&  in \prvkg & $n_\mathrm{prv}$ & $n_\mathrm{prv}^*$ & $n_\mathrm{match}$ & Precision & Recall & $n_\mathrm{prv}$ & $n_\mathrm{prv}^*$ & $n_\mathrm{match}$ & Precision & Recall & \\
\hline
\hline
CyberKG & \tfont{vulnerability} & 3,131 & 2,826 & 2,664 & 0.9427 & 0.8508 & 113,749 & 125,199 & 102,313 & 0.8172 & 0.8995 & 28,786\\
UMLS & \tfont{gene} & 17,475 & 15,128 & 12,435 & 0.8220 & 0.7116 & 638,616 & 510,951 & 440,900 & 0.8629 & 0.6904 & 249,071\\
YAGO & \tfont{therapy} & 1,727 & 1,946 & 1,557 & 0.8001 & 0.9016 & 116,458 & 84,483 & 75,325 & 0.8916 & 0.6468 & 49,597\\
\end{tabular}
\caption{Overall performance of \attack under the default  setting.\label{tab:expt_main}}}
\end{table*}

\begin{figure*}[!ht]
    \centering
    \epsfig{file = 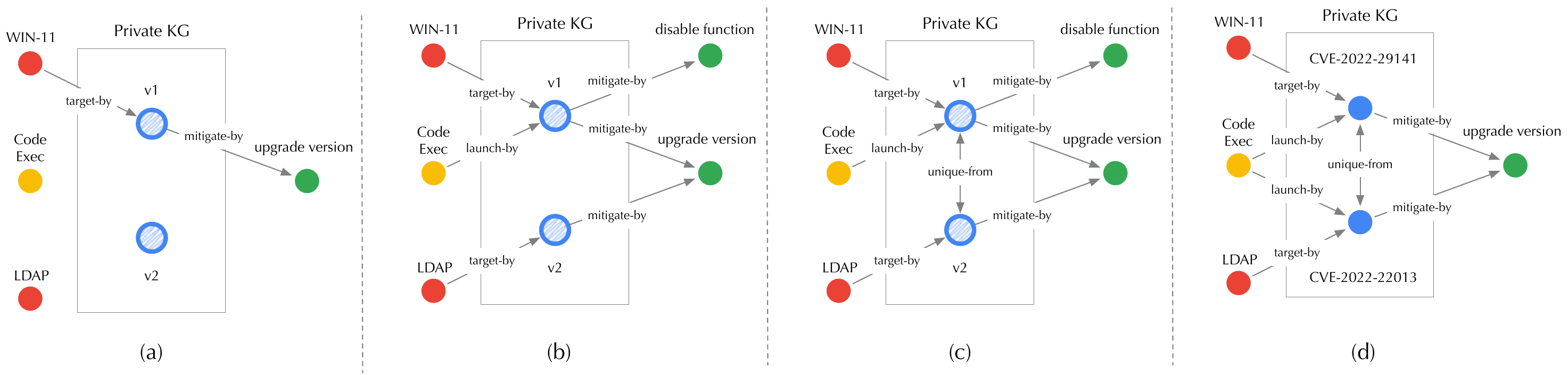, width =176mm}
    \caption{An illustrative case of \attack on CyberKG: (a)-(c) the extracted KG after using 100K/300K/500K queries; (d) target \kg.}
    \label{fig:running_example}
\end{figure*} 

\begin{figure*}[!ht]
    \centering
    \epsfig{file = 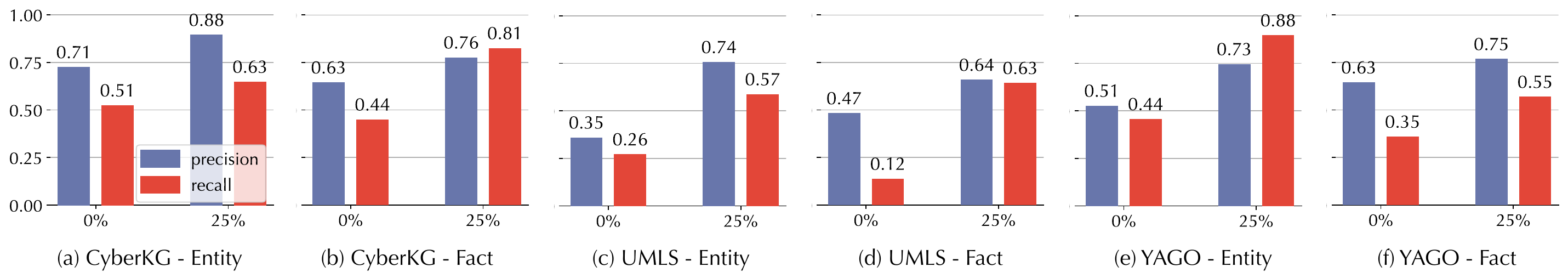, width =176mm}
    \caption{Performance of \attack under varying knowledge about \pubkg ($|\gG_\mathrm{pub}^*|/|\gG_\mathrm{pub}|$ = 0, 0.25).}
    \label{fig:pub_ratio}
\end{figure*}

\subsection{Evaluation results}
\label{ssec:eval}

\subsection*{Q1: KGX performance}

We first evaluate the overall performance of \attack.

\vspace{2pt}
{\bf Effectiveness --} Table\mref{tab:expt_main} summarizes the attack performance across different KGR APIs under the default setting. We have the following interesting observations.

\vspace{1pt}
\underline{{\em \attack is effective in KG extraction.}} Observe that \attack extracts relational structures with high resemblance to \prvkg across different KGs and KGR models (\meg, precision $\geq$ 0.80 and recall $\geq$ 0.64). As such relational structures capture the essential semantics of \prvkg, \attack serves as an effective KG extraction attack for completely reconstructing \prvkg.

\vspace{1pt}
\underline{\textit{\prvsur and \prvkg share a high overall similarity.}} Recall that graph edit distance (GED) is the minimum cost of transforming one graph to another. Consider the total number of facts in \prvsur and \prvkg: 238,948 for CyberKG, 1,149,567 for UMLS, and 200,941 for YAGO. The GED in Table\mref{tab:expt_main} accounts for only 12.05\%, 21.67\%, and 24.68\% of the total number of facts, respectively. We can thus conclude that beyond their matched parts $\gM$, \prvsur and \prvkg share a high overall similarity. Note that the absolute numbers of unmatched entities and facts are large-volume. In \msec{ssec:appendix_unmatch}, we provide a detailed analysis of properties on those unmatched entities and facts.


\vspace{2pt}
\underline{\textit{\kg properties matter.}} Observe that the measures of the same metric vary greatly across different \kgs. For instance, the recall of facts ranges from around 0.65 for YAGO to around 0.9 for CyberKG. This phenomenon may be explained by the complexity of the relational structures of different KGs. As shown in Table\mref{tab:kg}, CyberKG has 23 relation types, while YAGO has 78. It is relatively easier for \attack to infer the relational structures of CyberKG, compared with the more complex structures of YAGO. 

{\bf Interpretation on limited performance --}
Although the precision or recall are not astonishingly high, given the large size of backend KGs, the attack successfully extracts a large number of entities and relations under a limited query budget. For instance, 70\% recall on UMLS corresponds to 613K facts within total 870K facts in private sub-KGs. Moreover, due to the intricate relational structures, it is difficult to extract a perfect copy (e.g., around 100\% precision and recall) of \kgs, especially with only black-box accessibility. This work provides a proof-of-concept of how such attacks may be feasible in practical settings.

{\bf An illustrative case --} Figure\mref{fig:running_example} illustrates a running example of \attack on CyberKG, in which \attack gradually extracts the entity of \tfont{code-execution} relevant to \tfont{Windows}. Notably, with more queries, \attack extracts \prvsur with denser interconnections. However, Figure\mref{fig:running_example}\,(c) and (d) show the disparity of \prvsur and \prvkg: \prvsur misses the fact of \tfont{code-execution} $\sxrightarrow[]{\text{launch-by}}$ \tfont{CVE-2022-22013} while including facts non-existent in \prvkg (\meg, \tfont{CVE-2022-29141} $\sxrightarrow[]{\text{mitigate-by}}$ \tfont{disable-function}).

\subsection*{Q2: Influential factors}
\label{ssec:expt_factor}

Next, we evaluate the impact of external factors on \attack's performance. Specifically, we consider \mct{i} the knowledge about \pubkg, \mct{ii} the categories of private entities, \mct{iii} the connectivity of \prvkg, and \mct{iv} the query budget $n_\mathrm{query}$.

\vspace{2pt}
{\bf Knowledge about {\pubkg} --} Under the default setting, the adversary has access to a surrogate KG \surkg that comprises 50\% facts of \pubkg. Here, we evaluate the impact of the size of \surkg on \attack. Specifically, we consider the settings that \surkg comprises 25\% or 0\% (zero-knowledge) facts of \pubkg. Particularly, in the zero-knowledge case, the adversary has no access to \pubkg but only a set of seed entities in \pubkg (100 for CyberKG and YAGO, and 200 for UMLS). Figure\mref{fig:pub_ratio} shows the performance of \attack under these settings. We have the following observations.

\underline{{\em \attack retains effectiveness with less public knowledge.}} Observe that when \surkg comprises 25\% facts of \pubkg, \attack remains effective. For instance, the (entity) recall only degrades from 0.90 (\mcf Table\mref{tab:expt_main}) to 0.88 on YAGO. Under the zero-knowledge setting, \attack still achieves acceptable performance, especially on CyberKG and YAGO due to their relatively smaller sizes.

\begin{figure}[!ht]
    \centering
    \epsfig{file = 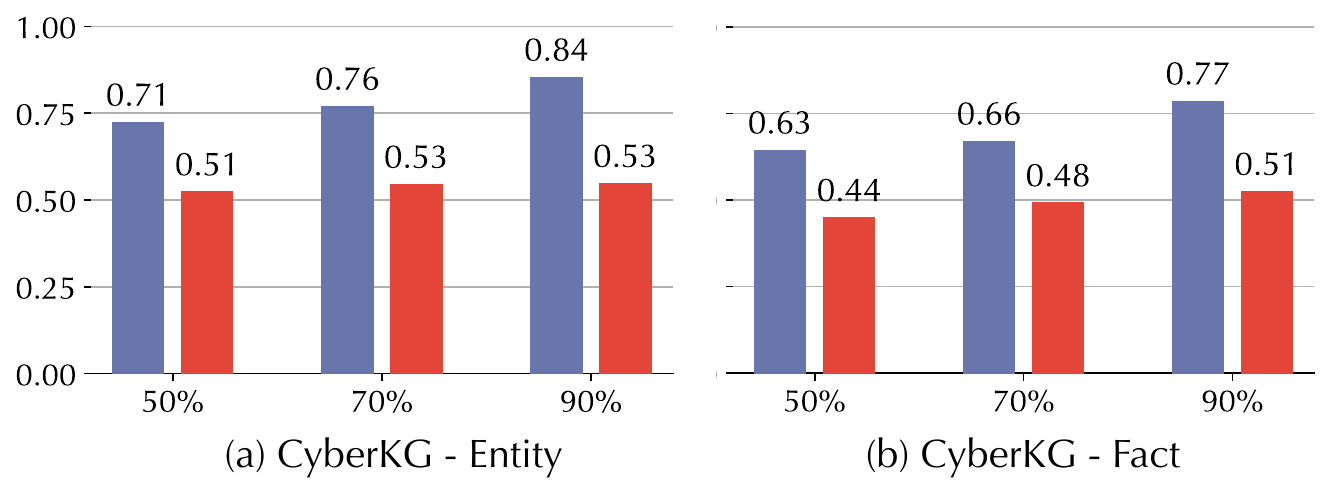, width=82mm}
    \caption{\attack performance under the zero-knowledge setting with the varying budget allocation (50\%, 70\%, 90\%) for \surkg expansion.}
    \label{fig:zero_pubratio}
\end{figure}

\underline{{\em Budget allocation impacts \attack's performance.}} Besides the knowledge about \pubkg, we point out that another factor that impacts \attack is the allocation of query budget $n_\mathrm{query}$. By default, we allocate 50\% of $n_\mathrm{query}$ to expand \surkg while using the remaining for KG extraction. Figure\mref{fig:zero_pubratio} evaluates the impact of this allocation on \attack's performance under the zero-knowledge setting. Observe that as the allocation for \surkg expansion increases from 50\% to 90\%, both precision and recall of \attack improve on CyberKG, suggesting the necessity of adjusting budget allocation when the knowledge about \pubkg is limited.  

\begin{table*}[!ht]{
\centering
\footnotesize
\setlength{\tabcolsep}{5pt}
\renewcommand{\arraystretch}{1.4}
\begin{tabular}{c|c|c|c|c|c|c|c|c|c|c|c|c}
 \multirow{2}{*}{\kg} & \multirow{1}{*}{Entity Category} & \multicolumn{5}{c|}{Entity} & \multicolumn{5}{c|}{Fact} &\multirow{2}{*}{GED}\\
 \cline{3-12}
 & in \prvkg  & $n_\mathrm{prv}$ & $n_\mathrm{prv}^*$ & $n_\mathrm{match}$ & Precision & Recall & $n_\mathrm{prv}$ & $n_\mathrm{prv}^*$ & $n_\mathrm{match}$ & Precision & Recall & \\
\hline
\hline
\multirow{2}{*}{CyberKG} & \tfont{mitigation} & 10,260 & 11,972 & 8,680 & 0.7250 & 0.8460 & 116,966 & 139,671 & 92,540 &  0.6626 & 0.7912 & 68,486 \\
& \tfont{attack pattern}  & 455 & 1,041 & 433 & 0.4159 & 0.9516 & 2,608 & 4,162 & 2,559 & 0.6148 & 0.9812 & 2,185 \\
\hline
\multirow{2}{*}{UMLS} & \tfont{lab procedure} & 33,025 & 35,746 & 24,761 & 0.6927 & 0.7498 & 869,615 & 753,015 & 612,614 & 0.8135 & 0.7045 & 394,118\\
& \tfont{protein} & 29,729 & 23,452 & 18,677 & 0.7964 & 0.6282 & 739,513 & 709,261 & 518,745 & 0.7314 & 0.7015 & 377,304\\
\hline
\multirow{2}{*}{YAGO} & \tfont{medical test} & 1,163 & 2,046 & 1,052 & 0.5141 & 0.9046 & 126,660 & 105,872 & 78,954 & 0.7457 & 0.6234 & 75,602\\
& \tfont{study and trail} & 872 & 1,905 & 826 & 0.4336 & 0.9472 & 132,360 & 110,734 & 91,864 & 0.8296 & 0.6940 & 58,460\\
\end{tabular}
\caption{\attack performance under the settings of alternative entity categories in \prvkg. \label{tab:prv_type}}}
\end{table*}

\begin{figure*}[!ht]
    \centering
    \epsfig{file = 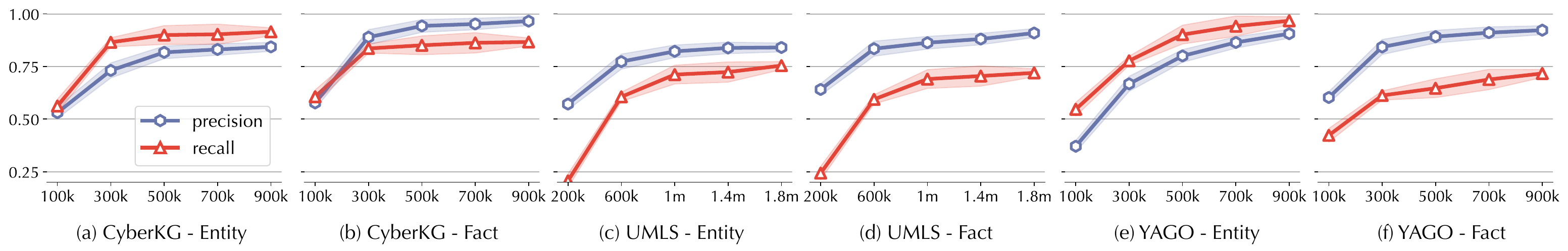, width =176mm}
    \caption{\attack performance under varying query budget $n_\mathrm{query}$.}
    \label{fig:cost}
\end{figure*} 

\begin{figure*}[!ht]
    \centering
    \epsfig{file = 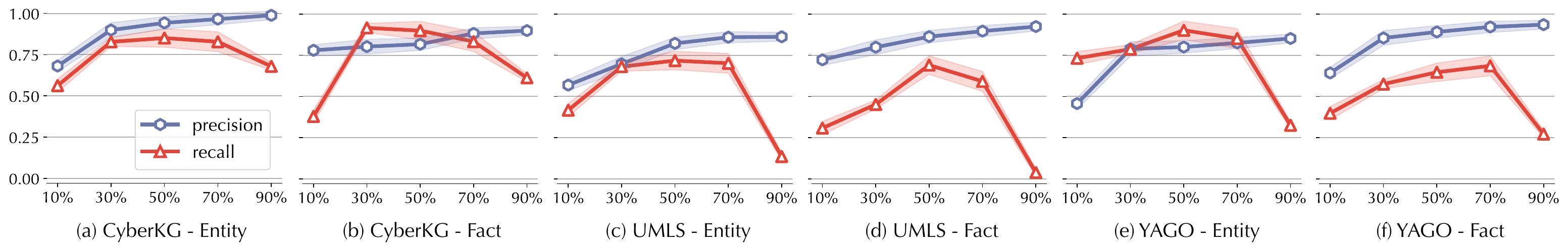, width =176mm}
    \caption{\attack performance under varying query budget allocated to expand surrogate KG \surkg.}
    \label{fig:pub_cost}
\end{figure*} 

\begin{figure*}[!ht]
    \centering
    \epsfig{file = 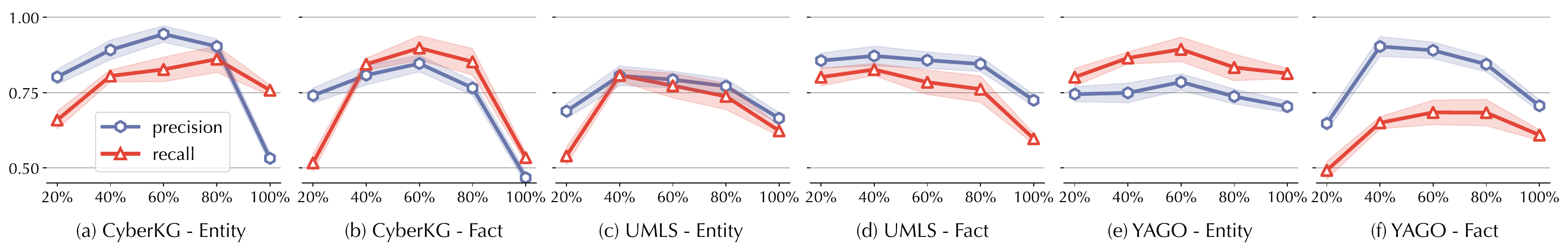, width =176mm}
    \caption{\attack performance under varying budget allocation for exploration and exploitation ($x$-axis as the fraction of exploration queries).}
    \label{fig:eps}
\end{figure*}

\vspace{2pt}
{\bf Entity categories in {\prvkg} --} By default, we assume the entities in \prvkg come from one fixed category (\mcf Table\mref{tab:expt_main}). Here, we consider that \prvkg consists of entities from alternative categories, with the remaining setting fixed. Table\mref{tab:prv_type} summarizes the evaluation results. We have the following observations.

\vspace{1pt}
\underline{{\em \attack is insensitive to \prvkg's entity categories.}} The entity category largely determines the size of \prvkg. For instance, if \prvkg comprises the category of \tfont{mitigation}, \prvkg includes 10,260 entities and 116,966 facts, while it includes 455 entities and 2,608 facts if it comprises the category of \tfont{attack pattern}. Despite this drastic variation, \attack achieves high precision and recall in extracting \prvkg across different settings. This may be explained by \attack's query generation strategies: regardless of the size of \prvkg, exploration and exploitation generate diverse queries that potentially intersect with \prvkg, which automatically adapt to large or small \prvkg.

\vspace{2pt}
{\bf Connectivity of {\prvkg} --} Thus far, we select \prvkg as a connected sub-KG from the target KG $\gG$. To evaluate the impact of \prvkg's connectivity on \attack, we consider the case that \prvkg consists of a set of disconnected sub-KGs. To simulate this case, we keep the same number and category of  entities as in Table\mref{tab:expt_main} but randomly sample entities and their facts from $\gG$ to form \prvkg. Table\mref{tab:rand_prv} shows the \attack performance, from which we have the following findings:

\begin{table}[!ht]{
\centering
\footnotesize
\setlength{\tabcolsep}{4pt}
\renewcommand{\arraystretch}{1.4}
\begin{tabular}{c|c|c|c|c|c|c}
 \multirow{2}{*}{Metric} & \multicolumn{2}{c|}{CyberKG} & \multicolumn{2}{c|}{UMLS} & \multicolumn{2}{c}{YAGO} \\
 \cline{2-7}
& Entity & Fact & Entity & Fact & Entity & Fact \\
\hline
\hline
$n_\mathrm{prv}$ & 3,131 & 181,659 & 17,475 & 286,961 & 1,727 & 62,058 \\
$n_\mathrm{prv}^*$ & 3,652 & 212,453 & 16,732 & 304,558 & 2,304 & 58,793 \\
$n_\mathrm{match}$ & 2,983 & 181,277 & 15,416 & 259,831 & 1,578 & 55,874 \\
Precision & 0.8168 & 0.8533 & 0.9213 & 0.8531 & 0.6849 & 0.9504 \\
Recall & 0.9527 & 0.9979 & 0.8822 & 0.9054 & 0.9137 & 0.9004 \\
\hline
GED & \multicolumn{2}{c|}{31,712} & \multicolumn{2}{c|}{71,013} & \multicolumn{2}{c}{9,578} \\
\end{tabular}
\caption{\attack performance on \prvkg with disconnected  structures.\label{tab:rand_prv}}}
\end{table}

\vspace{1pt}
\underline{{\em \attack remains effective to extract disconnected \prvkg.}} \attack retains its performance against disconnected \prvkg, as reflected in its high precision and recall. Moreover, compared with the connected \prvkg (\mcf  Table\mref{tab:expt_main}), the recall scores of both entities and facts increase. This may be intuitively explained as follows. The disconnected \prvkg consists of a set of sub-KGs with simpler structures. As \attack extracts \prvkg by probing its internal structures, simpler structures are easier to infer.

\vspace{2pt}
{\bf Query budget --} By default, we assume a fixed query budget $n_\mathrm{query}$ for the adversary. Here, we evaluate the impact of $n_\mathrm{query}$ on \attack's performance. Figure\mref{fig:cost} shows the results.

\vspace{1pt}
\underline{{\em There exists a ``diminishing gain'' effect.}} Although \attack's performance improves with $n_\mathrm{query}$, the performance gain decreases with $n_\mathrm{query}$. We may explain this phenomenon as follows. As more structures in \prvkg are uncovered, it becomes increasingly difficult for \attack to generate diverse and informative queries, slowing down its performance improvement. This also suggests dynamically adjusting the query generation strategies (\meg., focusing more on exploration rather than exploitation), which we consider as our ongoing research.

\subsection*{Q3: KGX practices} 

Next, we examine \attack's design spectrum and explore its  best practices under various settings.


\vspace{2pt}
{\bf Query budget allocation --} 
By default, we allocate half of the query budget $n_\mathrm{query}$ to expand  surrogate KG \surkg and the remaining for KG extraction. Here, we evaluate the impact of this budget allocation strategy on \attack's attack performance. Figure\mref{fig:pub_cost} summarizes the results, about which we have the following observations.

\vspace{1pt}
\underline{{\em Budget allocation balances precision and recall.}} Observe that both \attack's precision and recall steadily increase as more budget is allocated to expand \surkg. This is because richer knowledge about \pubkg improves the chance of generating high-quality queries to probe \prvkg, leading to more effective KG extraction. Meanwhile, if an excessive budget is allocated to expand \surkg (\meg, over 70\% for UMLS and YAGO), less is left to extract \prvkg, resulting in a significant decrease in \attack's recall. Therefore, we point out the necessity of properly allocating the budget to balance \attack's precision and recall (\meg, allocating 30-50\% on expanding \surkg for UMLS).

\begin{figure}[!ht]
    \centering
    \epsfig{file = 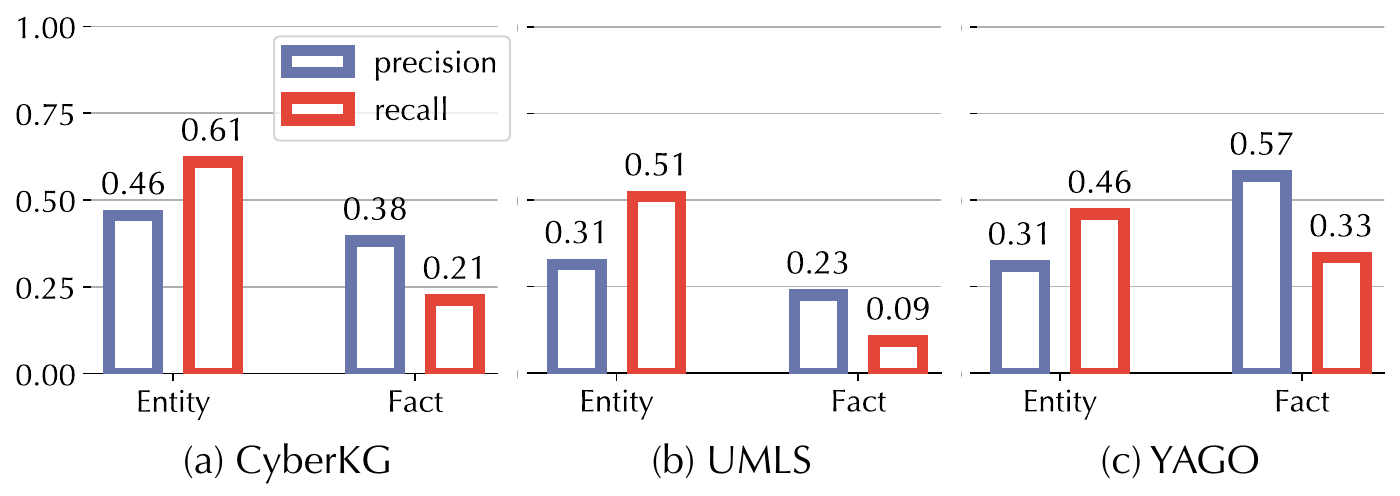, width =82mm}
    \caption{\attack performance without using any budget on expanding public knowledge.}
    \label{fig:no_pub_cost}
\end{figure} 

\vspace{1pt}
\underline{{\em Knowledge about \pubkg has a unique impact on \attack.}}
We further consider an extreme case that no budget is allocated to expand \surkg (\mie, all used for KG extraction). Recall that \surkg comprises 50\% of facts of \pubkg. Therefore, \attack is still able to attain reasonable performance, as shown in Figure\mref{fig:no_pub_cost}. However, there are large gaps between \attack's performance in Figure\mref{fig:pub_cost} and Figure\mref{fig:no_pub_cost}, especially for UMLS. This comparison indicates that obtaining sufficient knowledge about \pubkg is essential for \attack, especially for KGs with a large number of entities and facts (\meg, UMLS).

\vspace{2pt}
{\bf Exploration versus exploitation --} Recall that \attack adopts two query generation strategies, exploration and exploitation, both with limitations and strengths. We may thus optimize \attack by adjusting the query budget for both strategies. Here, we evaluate the impact of query budget allocation for exploration and exploitation on \attack's performance. Figure\mref{fig:eps} summarizes the results.

\vspace{1pt}
\underline{{\em There exists an exploration-exploitation trade-off.}} It is observed that across all the cases \attack attains the optimal performance under a proper trade-off between exploration and exploitation. Intuitively, only exploration lowers the chance of generating ``hit'' queries to probe \prvkg, while only exploitation refrains from effectively exploring unknown structures of \prvkg. Also, note that the optimal trade-off varies across different cases, which may be attributed to the unique characteristics of KGs and KGR models. We may empirically find the optimal exploration-exploitation trade-off.

\vspace{2pt}
{\bf Knowledge consolidation --} Besides query generation, another critical step in \attack is knowledge consolidation, which integrates the paths uncovered by individual queries to form more complex structures. We conduct an ablation study to investigate the importance of knowledge consolidation, with results shown in Table\mref{tab:no_consolid}.

\begin{table}[!ht]{
\centering
\footnotesize
\setlength{\tabcolsep}{3pt}
\renewcommand{\arraystretch}{1.4}
\begin{tabular}{c|c|c|c|c|c|c}
 \multirow{2}{*}{Metric} & \multicolumn{2}{c|}{CyberKG} & \multicolumn{2}{c|}{UMLS} & \multicolumn{2}{c}{YAGO} \\
 \cline{2-7}
& Entity & Fact & Entity & Fact & Entity & Fact \\
\hline
\hline
$n_\mathrm{prv}$ & 3,131 & 113,749 & 17,475 & 638,616 & 1,727 & 116,458 \\
$n_\mathrm{prv}^*$ & 94,490 & 320,765 & 463,453 & 1,703,868 & 226,358 & 451,119 \\
$n_\mathrm{match}$ & 2,664 & 102,313 & 13,976 & 520,600 & 1,572 & 85,236\\
Precision & 0.0282 & 0.3190 & 0.0301 & 0.3055 & 0.0069 & 0.1889 \\
Recall & 0.8508 & 0.8995 & 0.7998 & 0.8152 & 0.9102 & 0.7319 \\
\hline
GED & \multicolumn{2}{c|}{319,721} & \multicolumn{2}{c|}{1,729,925}  & \multicolumn{2}{c}{608,584} \\
\end{tabular}
\caption{\attack performance without knowledge consolidation.\label{tab:no_consolid}}}
\end{table}

\underline{{\em Knowledge consolidation is essential.}} Without knowledge consolidation, the extracted \kg \prvsur may contain a large number of entities and facts (\mie, $n_\mathrm{prv}^*$), wherein many entities (or facts) essentially correspond to the same entity (or fact) in the ground-truth \kg \prvkg. Thus, \attack's precision is significantly lower than its variant with consolidation (\mcf Table\mref{tab:expt_main}). Moreover, \prvsur comprises a set of relational paths rather than inter-connected structures, implying limited utility for the adversary. We thus highlight the necessity of applying knowledge consolidation in \attack to ensure the quality of the extracted KG \prvsur. 


\subsection*{Q4: Case Study: Google Knowledge Search API}
\label{ssec:case_study}

Next, using Google Knowledge Search API as a concrete case, we evaluate \attack against real-world KGR APIs. We first describe the experimental setting and then evaluate \attack under varying influential factors and practices.

\begin{figure*}[!ht]
    \centering
    \epsfig{file = 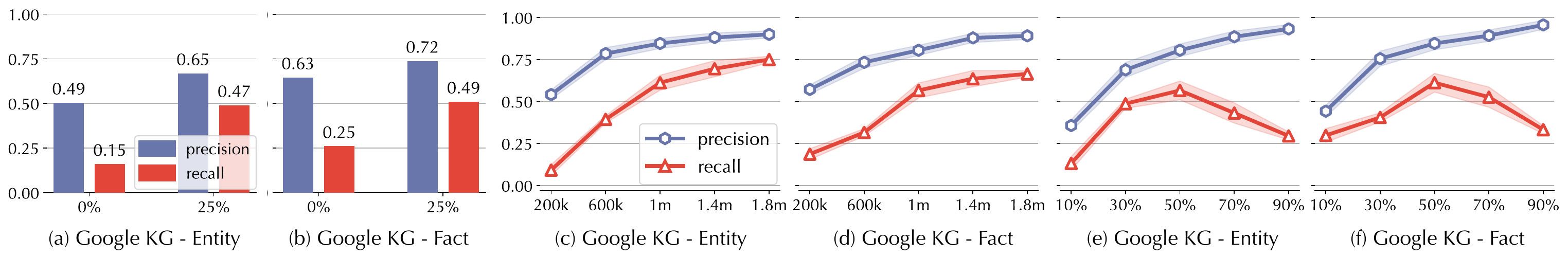, width =176mm}
    \caption{\attack performance on Google \kg under varying (a--b)  knowledge about \pubkg, (c--d) query budget $n_\mathrm{query}$, and (e--f) budget allocation to expand \surkg.}
    \label{fig:google_factors}
\end{figure*} 

\vspace{2pt}
{\bf Experimental setting --} As Google KG is not directly accessible to the public, we first build a ground-truth $\gG$ (for evaluation only) by iteratively querying the Google Knowledge Search API. Specifically, we start with two data sources of Google KG, Freebase \mcite{freebase-dump} and Wikidata \mcite{wikidata}. For each fact $s \sxrightarrow[]{r} o$ in Freebase or Wikidata with subject entity $s$, object entity $o$, and relation $r$, we generate a natural language query $q = (s, r)$ that combines the textual description of $s$ and $r$, and query the Google Knowledge Search API with $q$. 
For each entity $o' \in \llbracket q \rrbracket$ that is ranked higher than $o$, we add a new fact $s \sxrightarrow[]{r} o'$ to $\gG$. We consider $\gG$ as the ground-truth KG (for evaluation only), with its statistics summarized in Table\mref{tab:kg}.


We further consider \tfont{medical-topic} as the sensitive category and select its connected sub-KG as \prvkg (with the remaining of $\gG$ as \pubkg). Besides, we follow the default setting in Q$_1$--Q$_3$ in this case study, including \mct{i} 50\% facts of \pubkg as \surkg accessible to the adversary, \mct{ii} query budget $n_\mathrm{query}$ = 1,000,000; and \mct{iii} half of $n_\mathrm{query}$ to expand \surkg and the remaining half for KG extraction. The adversary aims to extract \pubkg by directly querying the Google Knowledge Search API. 

\vspace{2pt}
{\bf Experimental results --} We evaluate \attack under the default setting as well as varying influential factors and practices, which leads to the following observations.

 Further, we evaluate \attack under varying influential factors and practices, which are shown in   (ii)   (iii) Figure\mref{fig:google_factors} (c)(d) -- different amounts of available query budgets; (iv) Figure\mref{fig:google_factors} (e)(f) --ratio of budgets used to expand public knowledge. From those experiments, we have the following observations.

\begin{table}[!ht]{
\centering
\footnotesize
\setlength{\tabcolsep}{3pt}
\renewcommand{\arraystretch}{1.4}
\begin{tabular}{c|c|c|c|c|c|c}
& $n_\mathrm{prv}$ & $n_\mathrm{prv}^*$ & $n_\mathrm{match}$ & Precision & Recall & GED\\
\hline
 Entity & 45,024 & 31,659 & 25,475 & 0.8047 & 0.5658 &\multirow{2}{*}{517,908 } \\
 Fact & 1,122,663 & 812,453 & 686,732 & 0.8453 & 0.6117 & \\
\end{tabular}
\caption{\attack performance on Google \kg under the default setting.\label{tab:expt_google}}}
\end{table}

\underline{{\em \attack is effective to extract Google KG.}} Table\mref{tab:expt_google} summarizes \attack's performance on Google KG under the default setting. Recall that here we use the same query budget as in Q$_1$--Q$_3$, while Google KG is much larger than other KGs. Yet, \attack attains $\ge$ 0.80 precision and $\ge$ 0.56 recall on both entities and facts. Remarkably, \attack extracts much more ($\ge$ 13K) entities and ($\ge$ 24K) facts on Google \kg compared to UMLS (\mcf Table\mref{tab:expt_main}), using the same query budget. 

Additionally, Figure\mref{fig:google_factors}\,(a--b) shows \attack's performance under varying knowledge about \pubkg. Observe that \attack retains its effectiveness on Google \kg even with limited knowledge about \pubkg. For instance, it attains $\ge$ 0.65 precision and $\ge$ 0.47 recall when the initial \surkg consists of 25\% facts of \pubkg. Given the scale of Google \kg (\mie, around 3$\times$ entities and 2$\times$ facts of UMLS), \attack demonstrates its effectiveness on large KGs in real-world settings.

\begin{table}[!ht]{
\centering
\footnotesize
\setlength{\tabcolsep}{4pt}
\renewcommand{\arraystretch}{1.4}
\begin{tabular}{c|c|c|c|c|c|c}
& $n_\mathrm{prv}$ & $n_\mathrm{prv}^*$ & $n_\mathrm{match}$ & Precision & Recall & GED\\
\hline
 Entity & 45,024 & 36,541 & 28,880 & 0.7903 & 0.6414 &\multirow{2}{*}{159,427} \\
 Fact & 622,063 & 542,709 & 506,412 & 0.9331 & 0.8141 & \\
\end{tabular}
\caption{\attack performance on Google \kg, wherein \prvkg has scattered network structures.\label{tab:rand_prv_google}}}
\end{table}

\vspace{1pt}
\underline{{\em \attack performs better on disconnected \prvkg.}} Note that \prvkg is often disconnected in practice, due to the heterogeneous and distributed nature of real-world KGs. It is thus  crucial to evaluate the impact of \prvkg's connectivity on \attack. Similar to \msec{ssec:expt_factor}, we simulate the case that \prvkg consists of multiple, disconnected sub-KGs. As shown in Table\mref{tab:rand_prv_google}, \attack attains higher precision and recall when \prvkg consists of disconnected sub-KGs, compared with the case of connected \prvkg (\mcf Table\mref{tab:expt_google}). This observation corroborates the findings in \msec{ssec:expt_factor}. 

\underline{{\em Open-world KGs are more challenging.}} Although we use a pre-defined sub-KG \prvkg in evaluation, \attack queries the Google Knowledge Search API and interacts with the open-world Google KG. Despite \attack's effectiveness, it is worth noting that extracting open-world KGs is generally more challenging, due to their large scale and complexity, the adversary's limited prior knowledge and query budget, and the difficulty of optimizing attack strategies in black-box settings.   

Empirically, we observe: \mct{i} with zero-knowledge about \pubkg, it is more difficult for \attack to extract entities and facts than other \kgs, reflected in its lower recall in Figure\mref{fig:google_factors}\,(a--b) compared with UMLS in Figure\mref{fig:pub_ratio}\,(c--d); \mct{ii} the limited  query budget has a larger impact on \attack on Google KG, reflected in the substantial growth in recall with $n_\mathrm{query}$ in Figure\mref{fig:google_factors}\,(c--d); and \mct{iii} as shown in Figure\mref{fig:google_factors}\,(e--f), finding proper budget allocation to balance \surkg expansion and KG extraction is even more challenging, since both precision and recall vary more rapidly, compared with other KGs (\mcf Figure\mref{fig:pub_cost}). 

\section{Discussion}
\label{sec:discussion}


Declaring \prvkg is straightforward to block adversaries from acquiring confidential knowledge while easy to control by KGR developers (also defenders). However, as we have empirically shown in \msec{sec:expt}, \attack with constrained API accessibility and budgets effectively extracts the \prvkg, which implies the limitation of simply using \prvkg to protect privacy. Next, we study potential countermeasures on top of solely using \prvkg. Finally, we discuss the limitations of this work and outline several directions for future research.

\begin{figure*}[!ht]
    \centering
    \epsfig{file = 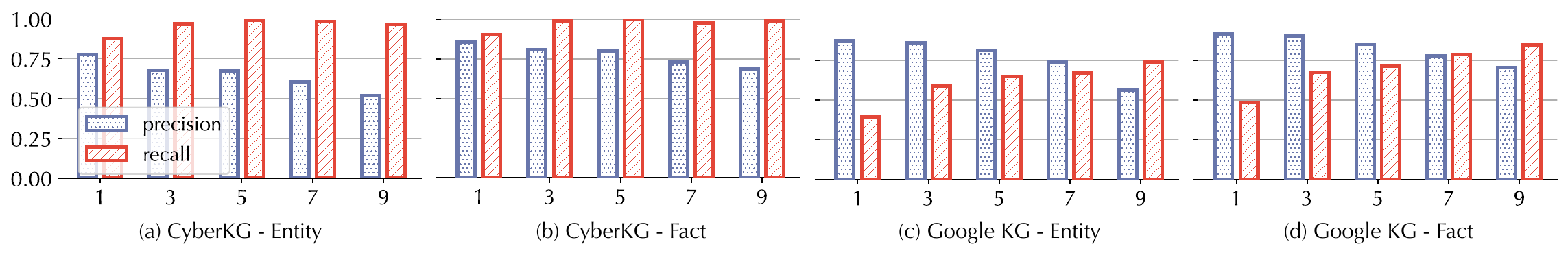, width =176mm}
    \caption{\system performance with respect to $k$ (only top-$k$ answers returned) on APIs with concealed confidence scores and shuffled rankings.}
    \label{fig:cm1}
\end{figure*} 





\subsection{Potential countermeasures}
\label{ssec:cm}

Due to the uniqueness of KGR tasks and KG extraction attacks, existing defenses against other privacy attacks \mcite{melis2019exploiting,he2019model} or in predictive tasks \mcite{pgd,neural-cleanse,spectral-signature} are often inapplicable. Instead, we investigate two potential countermeasures tailored to KGR APIs. Specifically, \mct{i} we may conceal the confidence scores of query answers and further shuffle their rankings to make it challenging for \attack to distinguish true and false answers; \mct{ii} motivated by defenses against model extraction attacks \mcite{tramer2016stealing}, we may also inject random noise into the confidence scores of query answers to mislead the adversary. We detail the countermeasures and findings as follows.

\vspace{2pt}
{\bf Concealing confidence scores --} Recall that the answer $\llbracket q \rrbracket$ of a given query $q$ is often a ranked list of entities with associated confidence scores. As \attack relies on the ranking and/or confidence information to distinguish true and false answers in probing the private sub-KG, a simple countermeasure is thus to conceal the confidence  and shuffle the ranking. Further, we may only return the top-$k$ entities in $\llbracket q \rrbracket$. Under this setting, we assume \attack validates path existence using bi-directional queries (details in \msec{ssec:query_gen}).

Figure\mref{fig:cm1} shows \system's performance on CyberKG and Google KG under this setting (with results on other \kgs deferred to Figure\mref{fig:cm1_appendix}). Observe that concealing confidence scores has a limited impact on \attack. With a sufficiently large $k$ (\meg, $k \geq 7$), \system attains fairly high recall (\meg, close to 1.0 in CyberKG) while maintaining acceptable precision (\meg, around 0.74 in CyberKG). This may be explained by that, even without confidence scores, bi-directional querying effectively validates path existence. Further, note that setting a small $k$ (\meg, $k=1$) is not always feasible. Despite its effectiveness against \attack, a small $k$ implies a high chance that the true answers are excluded from the query responses, which negatively impacts the utility of KGR APIs.  

\vspace{2pt}
{\bf Distorting confidence scores --} Alternatively, we consider distorting the confidence scores to mislead the adversary. We apply the Laplace mechanism from differential privacy to inject random noise into the confidence scores. Specifically, we sample noise from a Laplace distribution parameterized by $\epsilon$, with a smaller $\epsilon$ implying more significant noise.

\begin{figure*}[!ht]
    \centering
    \epsfig{file = 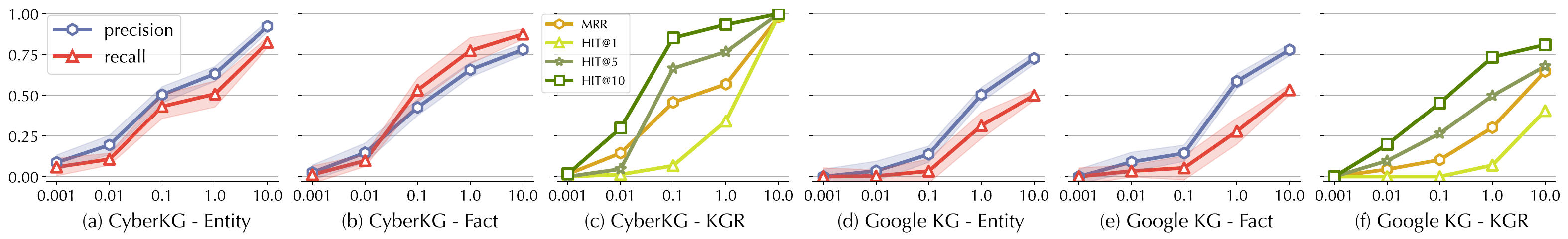, width =176mm}
    \caption{\system and \kgr performance on APIs with  confidence scores distorted by a Laplace distribution parameterized by $\epsilon$.}
    \label{fig:cm2}
\end{figure*} 

Figure\mref{fig:cm2} illustrates \attack's performance on CyberKG and Google \kg under varying $\epsilon$ (results on other KGs deferred to Figure\mref{fig:cm2_appendix}). We also assess the impact of this defense on the performance of KGR APIs, measured by the metrics of mean reciprocal rank (MRR) and HIT@$K$ ($K= 1, 5, 10$). Notably, a small $\epsilon$ (\meg, $\epsilon = 0.01$) effectively prevents \attack from extracting private sub-KGs. However, it also significantly impacts the KGR performance (\meg, MRR $\leq$ 0.2 on CyberKG). Apparently, there exists a delicate trade-off between attack robustness and KGR performance.

\subsection{Limitations and Future work}
\label{ssec:futurework}

We now discuss the limitations of this work and outline several directions for future research.

\vspace{2pt}
{\bf Complete KG extraction --} This work focuses on the critical and primary step of KG extraction, uncovering the relational structures of private sub-KG \prvkg. To completely re-construct \prvkg, we also need to map the nodes in the extracted sub-KG \prvsur to concrete entities. However, entity resolution typically requires auxiliary domain knowledge and resources. For instance, the adversary with threat-hunting expertise may explore threat reports to find specific relations between \tfont{vulnerability}, \tfont{software}, and \tfont{mitigation}. Then, by examining the relations between a \tfont{vulnerability} node in \prvsur and \tfont{software} and \tfont{mitigation} entities in \pubkg, the adversary may resolve this  \tfont{vulnerability} node. We consider developing principled methods for entity resolution as a promising research direction.

\vspace{2pt}
{\bf Alternative reasoning tasks --} This work mainly focuses on \kgr with individual entities as answers. There exist other reasoning tasks (\meg, path reasoning \mcite{path-reasoning} aims to find a logical path with a given pair of starting and ending entities). It is worth exploring the vulnerabilities of such alternative reasoning tasks to KG extraction attacks, and particularly, whether it is possible to extend \attack to such settings.

\vspace{2pt}
{\bf Defenses by design --} This work considers add-on defenses readily applicable to any KGR APIs without modification. Despite their simplicity, such defenses have apparent limitations. A more effective approach is to treat the robustness against KG extraction attacks as a first-class citizen and to build the attack robustness into the design of KGR systems.

\section{Related Work}
\label{sec:literature}

We survey relevant literature from three aspects including machine learning privacy, graph learning privacy, and knowledge graph privacy.

\vspace{2pt}
{\bf Machine learning privacy --} Many studies explore the privacy vulnerabilities of machine learning models \mcite{liu2021machine}. According to their objectives, the existing attacks can be roughly categorized as \mct{i} extracting information about training data \mcite{model-inversion,hitaj2017deep,melis2019exploiting,he2019model,salem2020updates}, \mct{ii} extracting model architectures and (hyper-)parameters \mcite{tramer2016stealing,wang2018stealing,orekondy2019knockoff,jagielski2020high}, and \mct{iii} various inference attacks (\meg, membership inference) \mcite{shokri2017membership,salem2018ml,nasr2019comprehensive,choquette2021label}. The current work follows this line of research and focuses on extracting confidential KGs via querying KGR APIs.

\vspace{2pt}
{\bf Graph learning privacy --} Motivated by the advances in graph neural networks (GNNs) \mcite{gcn, gat, graphsage}, one line of work specifically explores the privacy vulnerabilities of GNNs including \mct{i} quantifying privacy leakage in graph embeddings \mcite{duddu2020quantifying,ellers2019privacy,zhang2022inference}, \mct{ii} extracting GNN models \mcite{wu2022model,shen2022model}, and \mct{iii} inferring sensitive properties of training graphs (\meg, 
private links \mcite{he2021stealing,wu2022linkteller,zhang2021graphmi} and group properties \mcite{wang2022group}). 

This work differs from prior work in several major aspects:
\mct{i} adversary's objective -- \attack aims to uncover the relational structures of KGs while existing attacks focus on inferring specific links \mcite{he2021stealing}; \mct{ii}  adversary's knowledge -- \attack assumes the adversary only has black-box access to the KGR API, while prior attacks often assume additional knowledge about graphs (\meg, embeddings \mcite{ellers2019privacy}); and \mct{iii} target systems -- Rather than focusing on standalone models (\meg, GNNs), \attack targets KGR systems that integrate multiple models.

\vspace{2pt}
{\bf Knowledge graph privacy --} Despite the growing popularity of using KGs to support critical decision-making \mcite{gartner-report}, our understanding of the privacy and security risks of KGR is fairly limited. For instance, Xi {\em et al.} \mcite{kg-attack} explore the vulnerabilities of KGR to adversarial and poisoning attacks. To the best of our knowledge, this is the first study investigating the privacy vulnerabilities of KGR to KG extraction attacks. 
\section{Conclusion}

To sum up, this work represents the first study on the privacy risks of knowledge graph reasoning (\krl). We present \attack, a new class of KG extraction attacks that exploit black-box access to KGR APIs to reconstruct (``steal'') confidential KGs. We demonstrate the practicality of \attack under both experimental and real-world settings, raising concerns about the current practices of training and operating \krl. We also discuss potential countermeasures, which sheds light on applying KGR in a more principled and privacy-aware manner.

\clearpage

\bibliographystyle{plain}
\bibliography{reasoning}

\appendix

\section*{A. Details of KGR}
\label{sec:kgr_appendix}

Here, we detail the training and reasoning of \kgr.

{\bf Reasoning --} Representation-based \kgr typically consists of two main components:

\vspace{1pt}
\underline{Embedding function $\phi$} -- This function maps each entity in the knowledge graph $\gG$ as a latent embedding based on the relational structures within $\gG$. The entity embedding for entity $v$ is denoted as $\phi_v$, and the set of entity embeddings for all entities in $\gG$ is denoted as $\phi_\gG$.

\vspace{1pt}
\underline{Transformation function $\psi$} -- This function computes the embedding $\phi_q$ for a given query $q$. KGR defines a set of transformations, including the projection and intersection operators. The projection operator $\psi_r(\phi_v)$ computes the embeddings of entities connected to entity $v$ by relation $r$. The intersection operator $\psi_\wedge(\phi_{\gN_1}, \ldots, \phi_{\gN_n})$ computes the embeddings of entities that belong to the intersection of sets of entities $\gN_1, \ldots, \gN_n$. Other transformations, such as disjunction and negation, can be converted to the intersection operator\mcite{query2box, beta-embedding}. These transformation operators are often implemented as trainable neural networks\mcite{logic-query-embedding}.

To process query $q$, KGR starts with the query's anchors $\gA_q$ and iteratively applies the transformation function until reaching the entity of interest $v_?$, resulting in the query's embedding $\psi_q$. The entities in $\gG$ with embeddings most similar to $\psi_q$ are then identified as the query answer $\llbracket q \rrbracket$\mcite{bilinear-embedding}. 

\begin{figure}[!ht]
    \centering
    \epsfig{file = 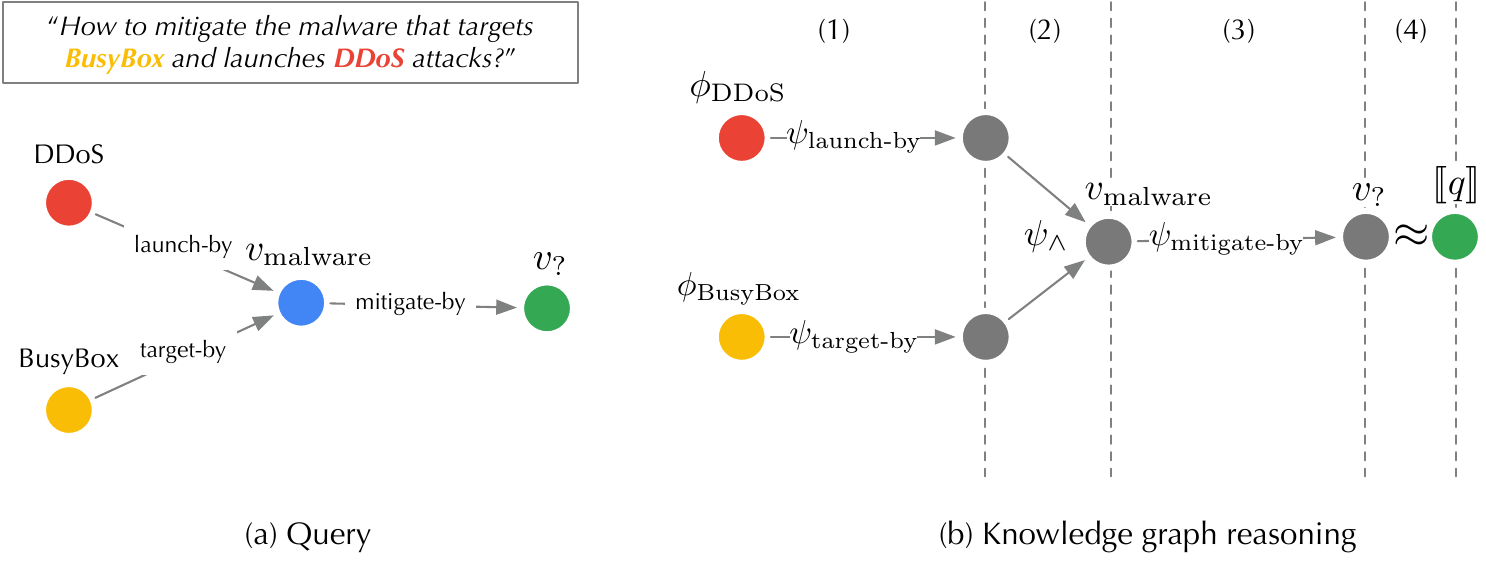, width =82mm}
    \caption{Illustration of \kgr reasoning.}
    \label{fig:kgr_appendix}
\end{figure}

\begin{example}
As shown in Figure\mref{fig:kgr_appendix}, \mct{1} starting from the anchors (\tfont{BusyBox} and \tfont{DDoS}), we apply the relation-specific projection operators to compute the entities connected to \tfont{BusyBox} and \tfont{DDoS} by the {\sl target-by} and {\sl launch-by} relations, respectively. \mct{2} We then use the intersection operator to identify the unknown variable $v_\text{malware}$. \mct{3} We further apply the projection operator to compute the entity $v_?$ that is connected to $v_\text{malware}$ by the {\sl mitigate-by} relation. \mct{4} Finally, we identify the entity most similar to $v_?$ as the answer $\llbracket q \rrbracket$.
\end{example}

{\bf Training --} \kgr is typically trained in a supervised manner. We sample a set of query-answer pairs $(q, \llbracket q \rrbracket)$ from $\gG$ as training data. The objective is to optimize the embedding and transformation functions by minimizing the prediction loss of each query $q$ with respect to the ground-truth answer $\llbracket q \rrbracket$.

\section*{B. Parameter Setting}
\label{sec:param_appendix}

Table\mref{tab:param} summarizes the default parameter setting in \msec{sec:expt}.

\begin{table}[!ht]{\footnotesize
\centering
\setlength{\tabcolsep}{3pt}
\renewcommand{\arraystretch}{1}
\begin{tabular}{c|r|l}
  & Parameter & Setting \\
\hline
\hline 
\multirow{8}{*}{\krl} & $\phi$ dimension & 200 \\
 & $\psi_r$ architecture & 4-layer FC\\
 & $\psi_\wedge$ architecture & 4-layer FC\\
 & Learning rate & 0.001 \\
& Batch size & 512 \\
& \krl training epochs & 50K \\
& \system optimization epochs & 10K \\
& Optimizer (\krl and \system) & Adam \\
\hline
\multirow{4}{*}{Attack} & \multirow{2}{*}{$n_\text{query}$} & 500K (CyberKG and YAGO) \\
& & 1,000K (UMLS and Google KG) \\
& Overlapping to \pubkg & 50\% \\
& Ratio of $n_\text{query}$ to extract \pubkg & 50\% \\
\hline
\end{tabular}
\caption{Default parameter setting.\label{tab:param}}}
\end{table}

\section*{C. Additional Results}
\label{sec:more_expt}

\subsection{Threshold of confidence scores}
\label{ssec:score}

\begin{figure}[!ht]
    \centering
    \epsfig{file = 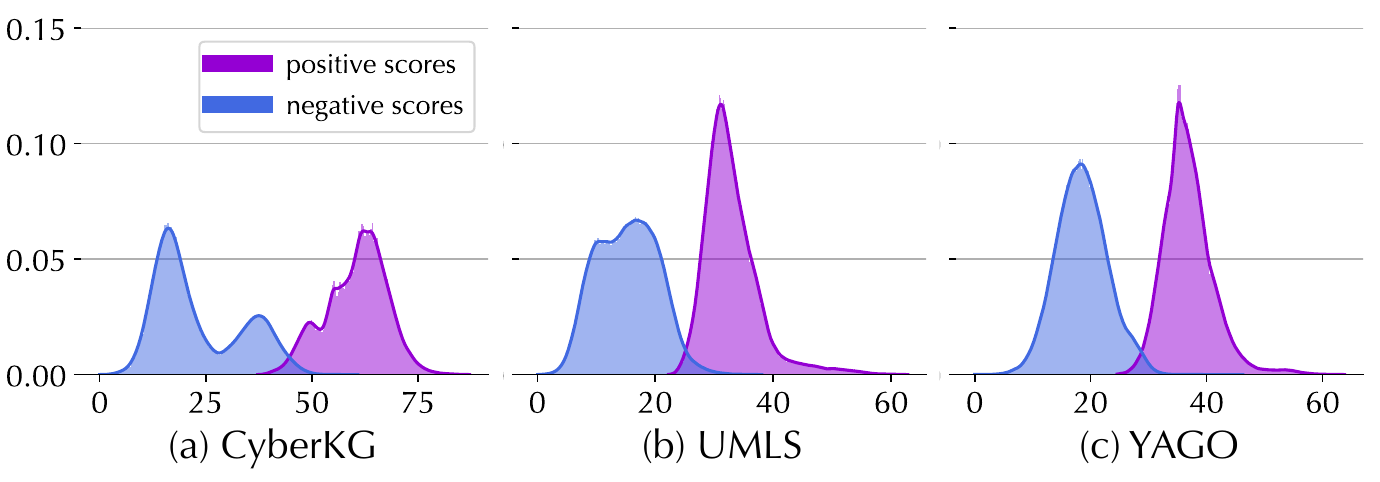, width =82mm}
    \caption{Distribution of confidence scores among positive and negative cases.}
    \label{fig:score}
\end{figure}

Here, we discuss how to empirically set the threshold $\lambda$ to distinguish true and false answers. Intuitively, we refer to  correct query answers as ``positive'' cases and denote their score distribution as $\gD^+$. Conversely, incorrect query answers are ``negative'' cases, and the distribution of their scores is $\gD^-$.

\vspace{1pt}
\underline{\em Are positive and negative cases separable?} We launch experiments to evaluate the disparity of $\gD^+$ and $\gD^-$. We sample 10K queries with various structures from different knowledge graphs such as CyberKG, UMLS, and YAGO. We collect
the scores of answers and non-answers as $\gD^+$ and $\gD^-$, respectively. In addition, we perturb the queries by introducing non-existent relations and use these perturbed queries to retrieve answer scores. We augment $\gD^-$ with the scores obtained from these perturbed queries. Figure\mref{fig:score} illustrates the distribution of $\gD^+$ and $\gD^-$. The areas under the curves are normalized to 1. Notably, $\gD^+$ and $\gD^-$ are mostly separate from each other, indicating the validity of using confidence scores to distinguish true and false answers.

\vspace{1pt}
\underline{\em How to set the threshold in \attack?} We first sample a small set of queries from the surrogate KG \surkg. Specifically, we sample 1,000 queries for CyberKG and YAGO, and 2,000 queries for UMLS and Google KG. We then retrieve the positive and negative scores ($\gD^+$ and $\gD^-$) using the same procedure as described earlier.
After obtaining the scores, we establish a threshold $\lambda$ that is higher than all the scores in $\gD^-$. This threshold filters all negative cases, even though at the cost of a few positive cases.

\begin{figure*}[!ht]
    \centering
    \epsfig{file = 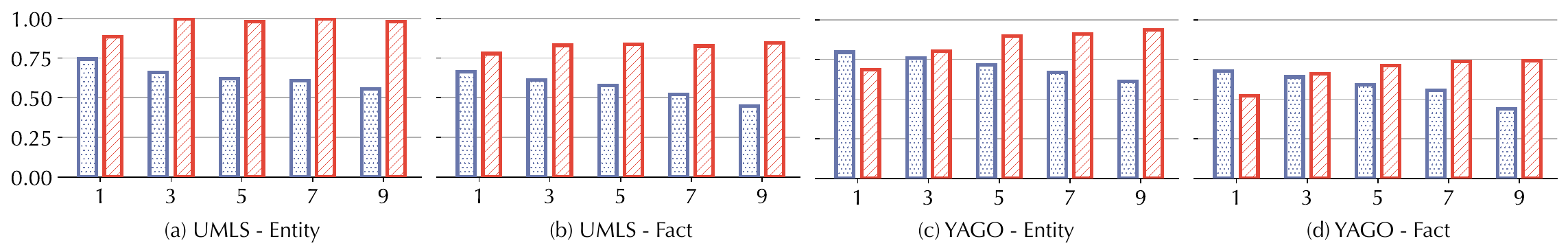, width =165mm}
    \caption{\system performance with respect to $k$ (only top-$k$ answers returned) on APIs with concealed confidence scores and shuffled rankings. Extending Figure\mref{fig:cm1}.}
    \label{fig:cm1_appendix}
\end{figure*} 

\begin{figure*}[!ht]
    \centering
    \epsfig{file = 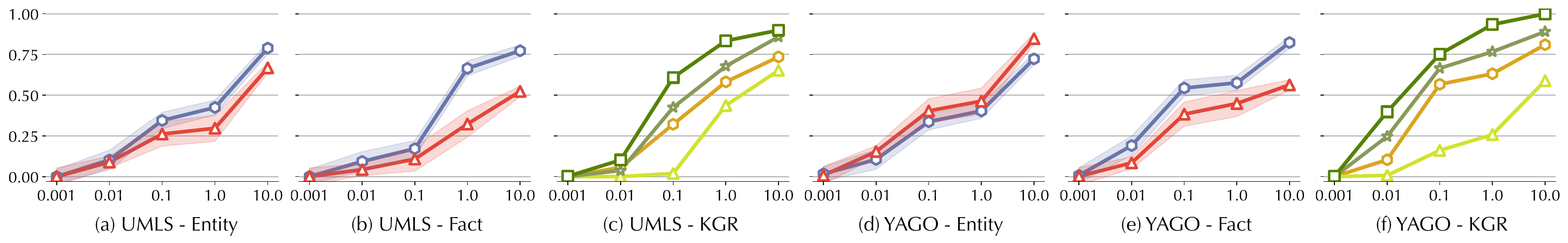, width =170mm}
    \caption{\system and \kgr performance on APIs with confidence scores distorted by a Laplace distribution parameterized by $\epsilon$. Extending Figure\mref{fig:cm2}.}
    \label{fig:cm2_appendix}
\end{figure*} 

\subsection{Unmatched entities/facts}
\label{ssec:appendix_unmatch}

\begin{table}[!ht]{\footnotesize
\centering
\setlength{\tabcolsep}{3pt}
\renewcommand{\arraystretch}{1.1}
\begin{tabular}{c|c|c|c}
 Min distance to \pubkg & \#matched & \#unmatched & unmatched ratio \\
\hline
1-hop & 9,181 & 3,654 & 39.8\% \\
2-hop & 3,208 & 1,353 & 42.2\% \\
3-hop & 46 & 33 & 71.7\% \\
\hline 
\end{tabular}
\caption{Number of matched and unmatched entities and unmatching ratios in UMLS -- \prvkg. Each row considers private entities with minimum 1/2/3-hop away from \pubkg}.\label{tab:match}}
\end{table}

We further analyze the properties of unmatched entities/facts to provide more insights for influential factors of successful graph matching. As shown in Table\mref{tab:match}, we group private entities in UMLS based on their minimum distance to \pubkg. With 1/2/3-hop distances, then measure ratios of unmatched entities. Notably, entities with larger distances to \pubkg are less likely to be matched. This implies that if entities in \prvkg are ``far away'' from \pubkg, \mie, the minimum distance to \pubkg is larger, they may be harder to be extracted and thus unable to be matched.

\subsection{Potential countermeasures}

Figure\mref{fig:cm1_appendix} and \mref{fig:cm2_appendix} show additional countermeasure results on UMLS and YAGO, extending Figure\mref{fig:cm1} and \mref{fig:cm2}, respectively.

\end{document}